\documentclass[aps,prl,twocolumn,superscriptaddress,reprint]{revtex4-2}

\usepackage{float}
\usepackage{graphicx}
\usepackage{amsmath}
\usepackage{bm}
\usepackage{color}
\usepackage{amssymb}
\usepackage{mathrsfs}
\usepackage{dcolumn}
\usepackage{multirow}
\usepackage{wasysym}
\usepackage[mathlines]{lineno}
\usepackage[colorlinks=true,linkcolor=blue,urlcolor=blue,citecolor=blue]{hyperref}

\begin{document}
	
	
	\title{Interaction-Driven Intervalley Coherence with Emergent Kekul\'e Orbitons}
	
	\author{Hua Chen}
	\email{Electronic address: hwachanphy@zjnu.edu.cn} 
	\affiliation{Department of Physics, Zhejiang Normal University, Jinhua 321004, China}

	\begin{abstract}		
	The $p$-orbital doublet in a honeycomb lattice is concretely studied with interacting spinless fermions at half filling. The Dirac fermions with linear dispersion at $\pm K$ valleys govern the non-interacting low-energy physics. In the weak-coupling regime, the Dirac fermions are gaped due to the spontaneous generation of mass terms through a uniform axial orbital ordering, rendering the system into the quantum anomalous Hall insulator phase with a nonzero Chern number. Surprisingly, the intermediate many-particle interaction produces the intervalley coherence between $\pm K$ valleys by developing complex polar orbital orderings in a tripled Wigner-Seitz cell. This phase is shown to have a deep connection with the low-energy physical behavior described by the orbital exchange model in the Mott insulating phase. The classical ground-state manifold in the Mott regime enjoys a continuous symmetry characterized by the intervalley coherent phase. Finally, the quantum fluctuation selects a unique ground state with emergent Kekul\'e orbitons through the order-by-disorder mechanism. Our findings provide insights for a direction of searching for Kekul\'e distortion in correlated multi-orbital systems. 	
	\end{abstract}
	
	\date{\today}
	
	\maketitle

	The recent experimental observations of Kekul\'e distortion in the graphene-based systems under a magnetic field~\cite{Li2019,Liu2022a,Coissard2022} and under a magic-angle twist~\cite{Nuckolls2023,Kim2023} have sparked immediate research interests towards the theoretical understanding about its mechanism. The observed Kekul\'e distortion in a $\sqrt{3}\times\sqrt{3}$ superlattice is elucidated as a salient feature of intervalley coherent insulator (IVCI) phase with the spontaneous translation-symmetry breaking, originating from the electronic correlations~\cite{Bultinck2020,Atteia2021,Kwan2021,Cifmmode2022,Hong2022,Christos2022,Das2022,You2022,Sanchez2024}. Generally, the intervalley coherent state
	\begin{equation}
		\left|\bm{\Omega}\left(\theta=\frac{\pi}{2},\varphi\right)\right\rangle
		=\cos\frac{\theta}{2}e^{-i\frac{\varphi}{2}}\left|+K\right\rangle
		+\sin\frac{\theta}{2}e^{i\frac{\varphi}{2}}\left|-K\right\rangle,
	\end{equation}
	arising from the many-particle scattering between the $\pm K$ valleys, is characterized by the azimuth angle $\varphi$ on the equator of Bloch sphere $\theta=\pi/2$. Despite the Kekul\'e distortion discussed in various contexts~\cite{Ajiki1995,Chamon2000,Hou2007,Frank2011,Wu2015,Gutierrez2016,Liu2017a,Gao2019,Menssen2020,Gao2020,Bao2021,Ma2021,Guan2024}, it is still highly desirable to explore fundamentally different mechanisms. Here, we present a comprehensive study of interacting spinless fermions in the $p$-band honeycomb lattice at half filling, exhibiting rich orbital orderings. In the weak-coupling regime, the Dirac fermions acquire a mass term through a uniform axial orbital ordering, resulting in the quantum anomalous Hall insulator phase with spontaneous time-reversal $\mathcal{T}$ symmetry breaking. Of particular interest is the IVCI phase with the complex polar orbital orderings in the $\sqrt{3}\times\sqrt{3}$ superlattice from the intermediate correlation regime. To unambiguously identify its nature, an effective orbital exchange model is constructed based on the Mott insulators (MIs) from the strong-coupling regime~\cite{Kugel1982,Brink1999,Brink2004,Nussinov2015}. We show that the classical ground-state manifold has a continuous symmetry dictated by the coherent phase $\varphi$ between $\pm K$ valleys with restored time-reversal $\mathcal{T}$ symmetry. The quantum fluctuation from orbitons finally selects a unique ground state by lifting the continuous degeneracy with emergent Kekul\'e orbitons. The order-by-disorder mechanism in supporting the Kekul\'e distortion can be distinguished from that of the graphene-based systems.

	\begin{figure}
		\centering
		\includegraphics[width=0.48\textwidth]{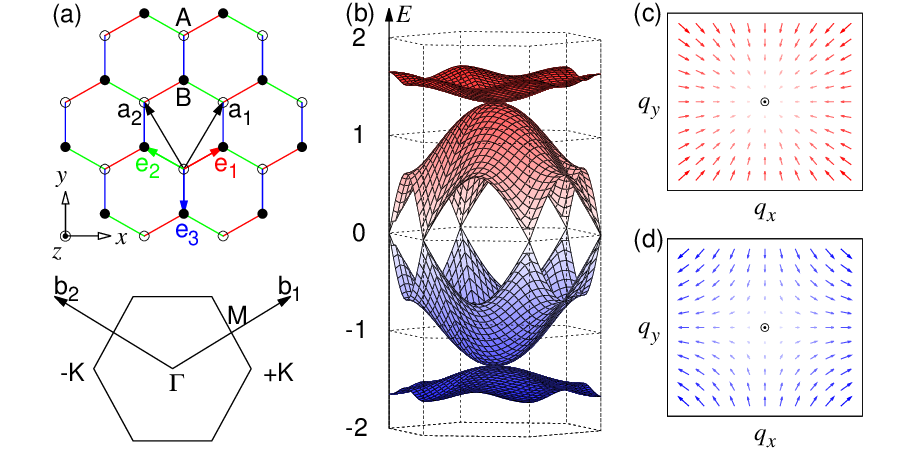}	
		\caption{
			(a) The bipartite structure of honeycomb lattice and the hexagonal Brillouin zone. 
			(b) The band structure of tight-binding model in Eq.~(\ref{eq:TB}) with $\{t_\sigma,t_\pi\}=\{1,-0.1\}$.
			The pseudovector field $\bm{p}\equiv\left(p_x,p_y\right)$ of the Dirac cones near (c) $+K$ and (d) $-K$ valleys resembles the vortex in XY systems with winding number $\mathcal{W}=\pm1$.
		}
		\label{fig:band}
	\end{figure}
	
	We start with the tight-binding model that describes the hopping process of spinless fermions in the $p$-band honeycomb lattice depicted in Fig.~\ref{fig:band}(a). Introducing an orbital-sublattice spinor representation $p_{\bm k}=\left[p_{x\mathcal{A}{\bm k}},p_{y\mathcal{A}{\bm k}},p_{x\mathcal{B}{\bm k}},p_{y\mathcal{B}{\bm k}}\right]^\text{T}$, the momentum-space Hamiltonian reads
	\begin{equation}
		H_{\text{TB}} = \sum_{\bm k} p^\dagger_{\bm k} \mathcal{H}_{\bm k} p_{\bm k},
		\mathcal{H}_{\bm k}=
		\left[
		\begin{matrix}
			0 & T_{\bm k} \\
			T^\dagger_{\bm k}  & 0
		\end{matrix}
		\right].
		\label{eq:TB}
	\end{equation}
	Here the elements of matrix $T_{\bm{k}}$ are given by
	\begin{equation}
		T^{\mu\nu}_{\bm{k}}=\sum_i
		\left[
		\left(t_\sigma-t_\pi\right)\hat{e}_i^\mu\hat{e}_i^\nu+\delta_{\mu,\nu}t_\pi
		\right]
		\epsilon_{i\bm{k}}.
	\end{equation}
	with the dispersion $\epsilon_{i\bm{k}}=\exp\left[i\bm{k}\cdot\left(\bm{e_i-e_3}\right)\right]$ describing the quantum tunneling from $p_\nu$ to $p_\mu$ orbital along $i$th bond vector $\bm{e}_i$.
	The hopping integral $t_\sigma$ and $t_\pi$ denote the $\sigma$ and $\pi$ bondings of $p$ orbitals, respectively. For the $\pi$ bonding, the bond vector lies in the nodal plane of $p$ orbitals. Consequently, the strength of $\pi$ bonding is typically weaker compared with the $\sigma$ bonding. The band structure with $\{t_\sigma,t_\pi\}=\{1,-0.1\}$, plotted in Fig.~\ref{fig:band}(b), can be qualitatively understood on the symmetry ground. Its symmetric feature with respect to zero energy arises from the chiral symmetry with the corresponding unitary operator $\mathcal{C}=s_z$. Here $s_z$ is the $z$-component Pauli matrix operating on the sublattice degree of freedom. Besides, the Bloch Hamiltonian in Eq.~(\ref{eq:TB}) enjoys the time-reversal symmetry $\mathcal{T}=\mathcal{K}$ with $\mathcal{K}$ being the complex conjugate operator. Finally, the particle-hole (charge-conjugation) symmetry is simply given by $\mathcal{P}=\mathcal{CT}$. The $\text{CPT}$ symmetries above complete the topological classification~\cite{Altland1997,Ryu2010}. The lower and upper two bands touch at $\Gamma$ point of hexagonal Brillouin zone (HBZ), and pin the Fermi level at filling $\nu=1/4$ and the particle-hole conjugated filling $\nu=3/4$, respectively. The low-energy physics is governed by the quadratic band touching points at $\Gamma$ point, which are generally unstable against many-particle interactions and lead to various topological phases~\cite{Wu2007,Chen2019}. While, the middle two bands cross at $\pm K$ valleys of HBZ. To describe the corresponding low-energy behavior, we introduce a pseudospin $\bm{\sigma}$ to label the doublet of degenerated zero-energy eigenstates $\psi_{\pm K}^{\sigma_z=+1}=\frac{1}{\sqrt{2}}\left[1,\pm i,0,0\right]^\text{T}$ and $\psi_{\pm K}^{\sigma_z=-1}=\frac{1}{\sqrt{2}}\left[0,0,1,\mp i\right]^\text{T}$ at $\pm K$ valleys, respectively. The effective two-band $k\cdot p$ Hamiltonian is given by 
	\begin{eqnarray}
		\mathcal{H}_{\pm K} \left({\bm q}\right) = 
		p_x\hat{\sigma}_x+p_y\hat{\sigma}_y+\mathcal{O}\left(q^2\right),
		\label{eq:kp}
	\end{eqnarray}
	where $\hat{\bm{\sigma}}$ are Pauli matrices and the coefficients
	\begin{eqnarray}
		\{p_x,p_y\} \equiv \frac{3}{4}&&\left(t_\sigma-t_\pi\right)\{\mp q_x,-q_y\}.
	\end{eqnarray}
	Diagonalizing $\mathcal{H}_{\pm K}\left({\bm q}\right)$ gives two non-interacting bands $E^\pm_{\pm K}\left({\bm q}\right)=\pm\sqrt{p_x^2+p_z^2}$, resulting in a linearly dispersed Dirac cone~\cite{Wu2008}. The pseudovector fields $\bm{p}\equiv\left(p_x,p_y\right)$ around $\pm K$ valleys shown in Figs.~\ref{fig:band}(c) and ~\ref{fig:band}(d) have $p$-wave symmetry. The topological charge of Dirac fermions is given by the winding number of pseudovector field: $\mathcal{W}=\frac{1}{2\pi}\oint_\mathcal{C}\nabla\theta\left({\bm q}\right)\cdot d{\bm q}=\pm 1$, where $\theta\equiv\text{arctan}\left(p_y/p_x\right)$ and $\mathcal{C}$ is a contour enclosing the singular Dirac node, indicating that the Driac fermions around $\pm K$ valleys carry a $\pm \pi$ Berry flux respectively~\cite{Volovik2009}. Hereafter we will focus on the half filling case.
		
	\begin{figure}
		\centering
		\includegraphics[width=0.48\textwidth]{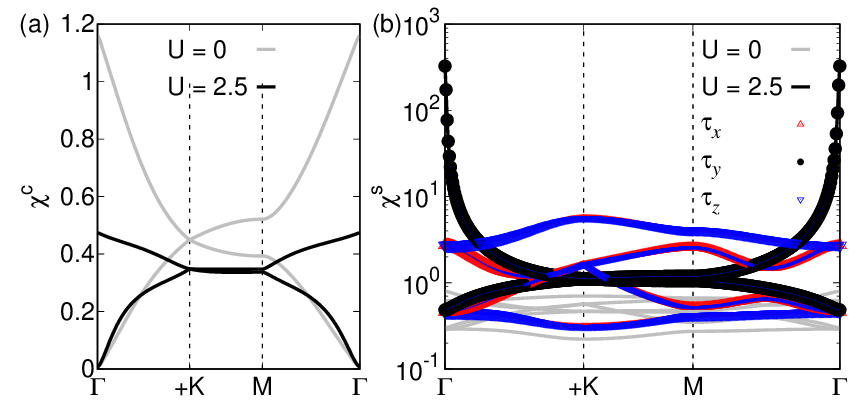}		
		\caption{ 
			Random-phase approximation for the static (a) charge $\chi^\text{c}$ and (b) pseudospin $\chi^\text{s}$ susceptibilities along the high-symmetry paths in HBZ for $U=0$ (gray solid lines) and $U=2.5$ (black solid lines) at $\{t_\sigma,t_\pi\}=\{1,-0.1\}$. The size of symbols in (b) for $U=2.5$ scales the projection of orbital pseudospin components by averaging over $\mathcal{A}$ and $\mathcal{B}$ sublattices. The inverse temperature $\beta=100$ is chosen.}
		\label{fig:chi}
	\end{figure}
		
	Having established the single-particle physics, we are next in a position to study the many-particle effects of Hubbard interactions $H_\text{I}=U\sum_i \hat{n}_{ix}\hat{n}_{iy}$ with density operator $\hat{n}_{i\mu}=p^\dagger_{i\mu}p_{i\mu}$. It is instructive to discuss the instabilities by adopting the standard random-phase approximation (RPA)~\cite{Pines2018}. Introducing a pseudospin $\tau_z=\pm1$ to label the $p_x$ and $p_y$ orbitals, the bare charge and pseudospin susceptibilities are defined as 
	\begin{eqnarray}
		\bar{\chi}^\text{c}_{\ell_1,\ell_2}(\bm{q},\omega)
		=\sum_{\{\mu\}}
		\tau^{\mu_1\mu_2}_0
		\bar{\chi}_{\mu_1\mu_2,\mu_3 \mu_4}^{\ell_1\ell_1,\ell_2 \ell_2}(\bm{q},\omega)
		\tau^{\mu_3 \mu_4}_0,	\\
		\bar{\chi}^\text{s}_{\ell_1\gamma_1,\ell_2\gamma_2}(\bm{q},\omega)
		=\sum_{\{\mu\}}
		\tau^{\mu_1\mu_2}_{\gamma_1}
		\bar{\chi}_{\mu_1\mu_2,\mu_3 \mu_4}^{\ell_1\ell_1,\ell_2 \ell_2}(\bm{q},\omega)
		\tau^{\mu_3 \mu_4}_{\gamma_2}
	\end{eqnarray}
	via the non-interacting susceptibility tensor
	\begin{eqnarray}
		\bar{\chi}_{\mu_1\mu_2,\mu_3 \mu_4}^{\ell_1\ell_2,\ell_3\ell_4}(\bm{q},\omega)
		=-\frac{1}{N}\sum_{\bm{k}}\sum_{ij}\frac{n_\text{FD}(\epsilon_{i\bm{k}})-n_\text{FD}(\epsilon_{j\bm{k}+\bm{q}})}{\hbar\omega+\epsilon_{i\bm{k}}-\epsilon_{j\bm{k}+\bm{q}}}	\nonumber\\
		\times\langle \ell_1\mu_1,i \bm{k}  | \ell_4\mu_4, i \bm{k}  \rangle	
		\langle \ell_3\mu_3,j \bm{k}+\bm{q} | \ell_2\mu_2, j \bm{k}+\bm{q} \rangle.	
		\hspace{3mm}
		\label{eq:chi0}
	\end{eqnarray}
	Here the indexes $\ell=\{\mathcal{A},\mathcal{B}\}$ and $\mu=\{p_x,p_y\}$ with subscripts label the sublattice and orbital degrees of freedom, respectively, the Fermi-Dirac distribution function $n_\text{FD}(\epsilon)=1/\left(\exp\left[\beta\epsilon\right]+1\right)$ and the Dirac notions in Eq.~(\ref{eq:chi0}) represent the Bloch wave functions of the Hamiltonian $\mathcal{H}_{\bm{k}}$ in Eq.~(\ref{eq:TB}). The charge and pseudospin susceptibilities under RPA are calculated by
	\begin{equation}
		\chi^{\text{c/s}}(\bm{q},\omega)
		=\left[1+\bar{\chi}^{\text{c/s}}(\bm{q},\omega)\Gamma^{\text{c/s}}\right]^{-1}
		\bar{\chi}^{\text{c/s}}(\bm{q},\omega)
		\label{eq:chi}
	\end{equation}
	with the matrix elements of the charge- and pseudospin-channel interaction vertex being parameterized as $\Gamma^\text{c}_{\ell,\ell^\prime}=\frac{U}{2}\delta_{\ell,\ell^\prime}$ and $\Gamma^\text{s}_{\ell\gamma,\ell^\prime\gamma^\prime}=-\frac{U}{2}\delta_{\ell,\ell^\prime}\delta_{\gamma\gamma^\prime}$, respectively. Figures~\ref{fig:chi}(a) and \ref{fig:chi}(b) plot the eigenvalues of static charge $\chi^\text{c}$ and pseudospin $\chi^\text{s}$ susceptibilities along the high-symmetry paths in HBZ, respectively. The charge channel $\chi^\text{c}(\bm{q},\omega=0)$ is suppressed by switching on the many-particle interaction $U$. In contrast, the pseudospin fluctuations $\chi^\text{s}(\bm{q},\omega=0)$ are significantly enhanced. The orbital ordering can be further inferred from the eigenvectors of pseudospin susceptibility $\chi^\text{s}$. The most pronounced peak at $\Gamma$ point suggests that the leading instability towards orbital pseudospin ordering along $y$ axis originates from the intravalley scattering. Moreover, the onset of orbital ordering at the critical interaction $U_c\approx2.51$ can be predicted through the generalized Stoner criteria $\det\left[1+\chi_0^{\text{s}}(\bm{q}=\Gamma,\omega=0)\Gamma^{\text{s}}\right]=0$, which indicates the divergence of pseudospin susceptibility $\chi^\text{s}$. Besides, the broad peak at $+K$ point of HBZ, derived from the intervalley scattering, is dominated by the orbital ordering in $zx$ plane of pseudospin space. 
	
	\begin{figure}
		\centering
		\includegraphics[width=0.48\textwidth]{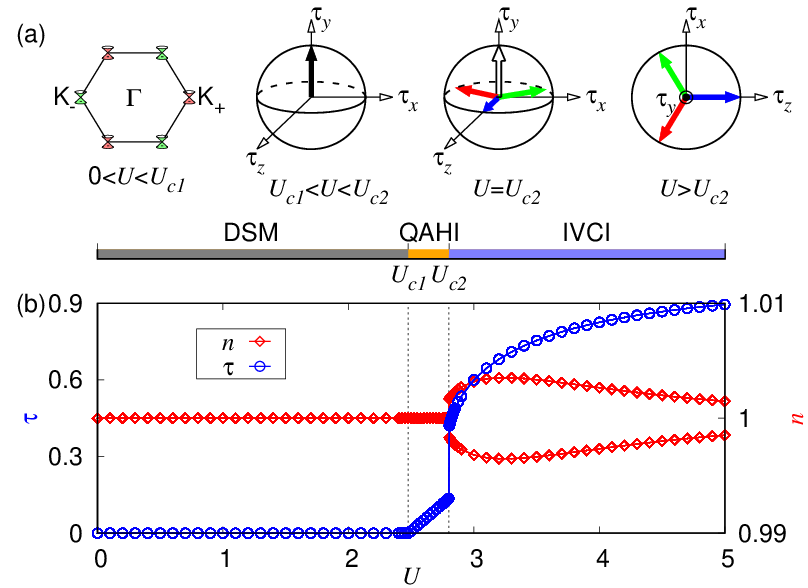}		
		\caption{
			(a) The phase diagram as a function of Hubbard interaction $U$ at $\{t_\sigma,t_\pi\}=\{1,-0.1\}$ shows three phases:
			(1) Dirac semimetal (DSM) with linearly dispersed Dirac fermions at $\pm K$ valleys.
			(2) Quantum anomalous Hall insulator (QAHI) with the spontaneous $y$-axis pseudospin ordering.
			(3) Intervalley coherent insulator (IVCI) with the pseudospin ordering aligned in $zx$ plane and the nonuniform charge orderings on sublattice $\mathcal{A}$ and $\mathcal{B}$.
			(b) The evolution of pseudospin magnitudes $\tau$ and charge $n$ orderings. 
			The black dashed line at $U_{c1}$ marks the continuous transition that separates DSMs from QAHIs.
			The black dashed line at $U_{c2}$ marks the discontinuous pseudospin-flop  transition between QAHIs and IVCIs.
		}
		\label{fig:hf}
	\end{figure}
	
\begin{figure}
		\centering
		\includegraphics[width=0.48\textwidth]{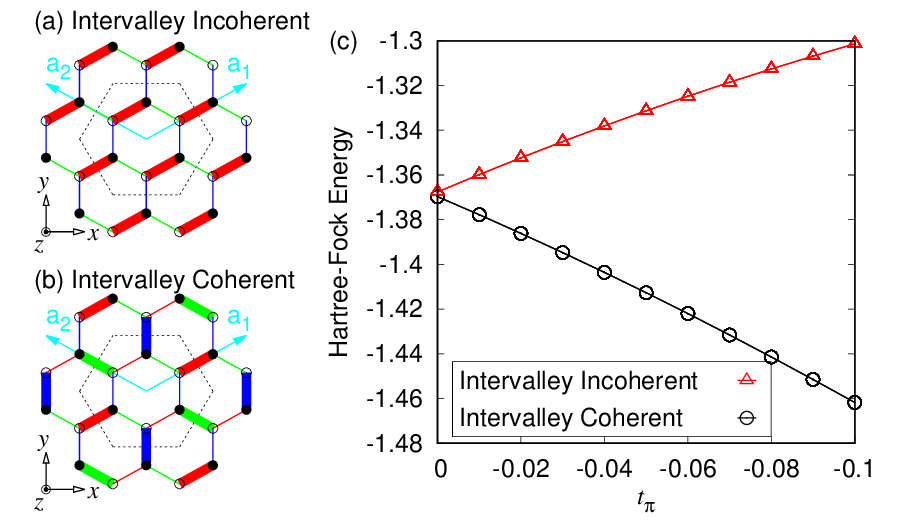}		
		\caption{
			The pictorial representations of pseudospin textures for the intervalley (a) incoherent and (b) coherent phases allowed in the tripled Wigner-Seitz cell. The bonds aligned with anti-parallel pseudospins are highlighted by thick lines. (c) The ground-state energies as a function of $t_\pi$ for both the intervalley incoherent and coherent phases at $\{t_\sigma,U\}=\{1,5\}$.
		}
		\label{fig:dimer}
	\end{figure}

	The charge and pseudospin susceptibilities within RPA offer an unbiased prediction on the instability towards various orderings but become invalid beyond the critical interaction $U>U_c$. Therefore, the Hartree-Fock approximation is further implemented to establish the phase diagram by taking both the intra- and inter-valley scattering processes into account. The latter process folds $\pm K$ valleys to $\Gamma$ point, accompanied by tripling the Wigner-Seitz cell of honeycomb lattices. The Hubbard interaction, at mean-field level, is described by Hartree and pseudospin exchange self-energies
	\begin{equation}
		H_\text{I}
		=\frac{U}{2}\sum_i\left(n_i\hat{n}_i-\bm{\tau}_i\cdot\hat{\bm{\tau}}_i-\frac{n_i^2-\bm{\tau}_i^2}{2}\right)
		\label{eq:HF}
	\end{equation}
	where $\hat{n}_i=\sum_{\mu=x,y}p^\dagger_{i\mu}p_{i\mu}$ and $\hat{\bm{\tau}}_i=\sum_{\mu\nu}p^\dagger_{i\mu}\bm{\sigma}_{\mu\nu}p_{i\nu}$ are the total density and pseudospin operators at the $i$-th site, with $n_i$ and $\bm{\tau}_i$ as the ground-state expectation values. The first term in Eq.~(\ref{eq:HF}) renormalizes the on-site energy level. While, the pseudospin exchange interaction, the second term in Eq.~(\ref{eq:HF}), preserves the pseudospin rotational symmetry of the Hubbard interaction and favors orbital orders by lowering the exchange self-energy. As sketched in Fig.~\ref{fig:hf}(a), the calculated phase diagram with the hopping integrals $\{t_\sigma,t_\pi\}=\{1,-0.1\}$ accommodates three different phases including DSMs, QAHIs, and IVCIs. Figure~\ref{fig:hf}(b) plots the evolution of charge $n$ and pseudospin magnitude $\tau$ as a function of Hubbard interaction $U$. Across the entire phase diagram, all sublattices in the tripled Wigner-Seitz cell develop an identical magnitude of pseudospin ordering. Initially, a finite Hubbard interaction $U_{c1}<U<U_{c2}$ produces a $y$-axis pseudospin ordering on top of DSMs, as a result of spontaneous time-reversal symmetry $\mathcal{T}$ breaking. Consequently, the Dirac fermion at $\pm K$ valleys acquires a mass term $\mp \frac{U}{2} \tau_y \hat{\sigma}_z$ in addition to the effective $k\cdot p$ Hamiltonian in Eq.~(\ref{eq:kp}) respectively, accompanied by the opening of a topological band gap. This insulating phase has non-trivial band topology characterized by the Chern number
	\begin{eqnarray}
		\text{Ch}&=&\int_\text{HBZ}d^2{\bm k}
		\sum_{m\ne n}\left[n_\text{FD}\left(E^n_{\bm k}\right)
		-n_\text{FD}\left(E^m_{\bm k}\right)\right] \nonumber\\
		&&\times \frac{1}{2\pi} 
		\frac{\text{Im}\left[	\langle n{\bm k}|\hat{v}_x| m{\bm k}\rangle 
			\langle m{\bm k}|\hat{v}_y| n{\bm k}\rangle \right]}
		{\left(E^n_{\bm k}-E^m_{\bm k}\right)^2}
		\label{eq:chern}
	\end{eqnarray} 
	with $\hat{v}_{\mu}=\partial\mathcal{H}_{\bm k}/\partial k_\mu$ being the velocity operator. Explicit evaluations of Eq.~(\ref{eq:chern}) give $\text{Ch}=\pm1$ as the result of spontaneous time-reversal symmetry $\mathcal{T}$ breaking by freely selecting the pseudospin ordering aligned along $\mp y$ axis. The nontrivial topological property arises from the orbital angular momentum of axial $p_\pm=p_x \pm ip_y$ orbital orderings with the orbital degeneracy lifted by the pseudospin exchange along $\mp y$ axis. In the insulating case, the Hall conductance is determined by the Chern number of the occupied bands and must be an exact integer in units of the conductance quantum $e^2/h$~\cite{Klitzing1980,Thouless1982,Kohmoto1985}. Correspondingly, the system has a quantized Hall conductivity $\sigma_{xy}=\pm e^2/h$ in the absence of external magnetic fields and supports a single gapless chiral edge mode in each edge channel. This phase is thus identified as QAHIs. Upon increasing the Hubbard interaction $U$, the nonuniform charge ordering distributed on sublattice $\mathcal{A}$ and $\mathcal{B}$ sets in at the critical Hubbard interaction $U_{c2}$ but is gradually suppressed due to the frozen charge fluctuations in the large $U$ limit. Meanwhile, a pseudospin-flop transition from $y$ axis to $zx$ plane occurs with the discrete bond-locked pseudospin orientations ${\bm \tau}$ depicted in Fig.~\ref{fig:hf}(a), recovering time-reversal symmetry $\mathcal{T}$. This phase is consistent with the RPA prediction that the broad peak of pseudospin susceptibility at $+K$ point arises from the intervalley scattering process. To gain overall understanding on the effects of $\pi$ bonding $t_\pi$, we evaluate the ground-state energies for both the intervalley incoherent and coherent states as depicted in Figs.~\ref{fig:dimer}(a) and \ref{fig:dimer}(b), respectively. Figure~\ref{fig:dimer}(c) shows that the $\pi$ bonding $t_\pi$ favors the intervalley coherent phase with a lower ground-state energy. The pseudospins ${\bm \tau}$ in Fig.~\ref{fig:dimer}(b) have a 120$^\circ$ N\'eel order on one set of sublattice $\mathcal{A}$ and $\mathcal{B}$ as indicated by the corresponding colors in Fig.~\ref{fig:hf}(a). From symmetry aspects, the translation symmetry breaking of honeycomb lattice shown in Fig.~\ref{fig:dimer}(b) is simply a manifestation of the intervalley coherence. The phase transition between QAHIs and IVCIs is expected to be of first-order type due to distinct broken symmetries of these two phases. The corresponding phase boundary is thus indicated by the discontinuous jumps of both the charge $n$ and pseudospin magnitude $\tau$, reflecting an abrupt pseudospin-flop transition. 
	
	\begin{figure}[b]
		\centering
		\includegraphics[width=0.48\textwidth]{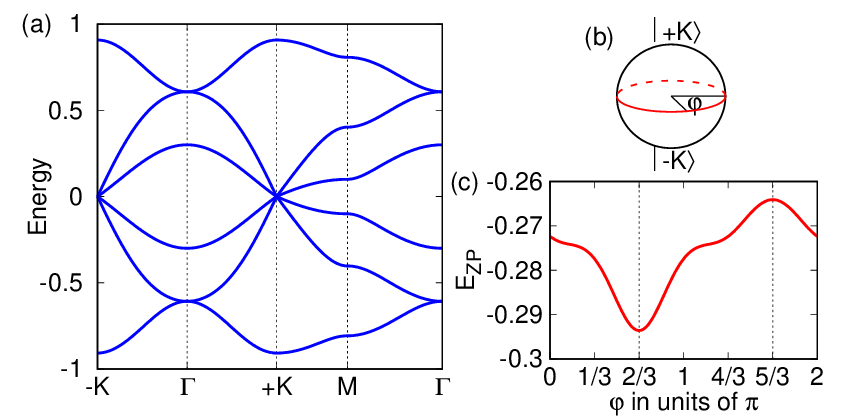}	
		\caption{ 
			(a) The eigenvalues of orbital interaction matrix $\Lambda_{\bm{k}}$ within the classical approximation.
			(b) Bloch sphere with the north and south poles corresponding to $+K$ and $-K$ valleys, respectively.
			(c) The zero-point energy $E_\text{ZP}\left(\varphi\right)$ in Eq.~(\ref{eq:zp}) from the correction of quantum fluctuations has a minimum at $\varphi=2\pi/3$. The energies in (a) and (c) are in units of $1/U$ with $\{t_\sigma,t_\pi\}=\{1,-0.1\}$.
		}
		\label{fig:sw}
	\end{figure}
	
	For sufficiently large $U$, the system is essentially MIs rather than Slater insulators, which cannot be captured by the Hartree-Fock approximation. Since the charge excitation is frozen by the Mott Hubbard gap, the orbital fluctuation is the remaining low-energy degree of freedom in MI regime. Therefore, the superexchange model can be constructed with the virtual hopping processes being treated perturbatively. Following the standard second-order perturbation theory~\cite{MacDonald1988,Fazekas1999,Kuklov2003}, the effective low-energy physics for the MI regime is captured by the following orbital exchange model
	\begin{eqnarray}
		H_\text{OE}=J\sum_{\langle ij \rangle\parallel\bm{e}_\gamma}\tau^i_\gamma\tau^j_\gamma
		+J^\prime \sum_{\langle ij \rangle} \bm{\tau}^i\cdot\bm{\tau}^j
		\label{eq:orbex}
	\end{eqnarray}
	with
	\begin{eqnarray}
		\tau_\gamma=\tau_z\cos\left[2\theta_\gamma\right]+\tau_x\sin\left[2\theta_\gamma\right] 
		\text{ and }
		\theta_{\gamma} = -\frac{1}{6}\pi+\frac{2\gamma}{3}\pi.\nonumber
	\end{eqnarray}
	Detailed derivations are presented in Supplemental Material. The antiferro-orbital exchange $J=(t_\sigma-t_\pi)^2/2U$ in Eq.~(\ref{eq:orbex}) is inherently anisotropic originating from the orientation of $p$ orbitals. In contrast, the isotropic exchange $J^\prime=t_\sigma t_\pi/U$ is ferro-orbital due to the opposite sign of $t_\sigma$ and $t_\pi$. The first term in Eq.~(\ref{eq:orbex}) involving interacting orbital degree of freedom is coined as the compass model~\cite{Kugel1982,Brink2004,Nussinov2015}. The ground state is first studied by treating the orbital pseudospin $\bm{\tau}$ as a classical vector based on the Luttinger-Tisza approach~\cite{Luttinger1946}. The orbital interaction can be minimized via the diagonalization of the orbital exchange Hamiltonian in momentum space $H_\text{OE} = \sum_{\bm{k}}\bm{\tau}^{-\bm{k}}\Lambda_{\bm{k}}\bm{\tau}^{\bm{k}}$. As shown in Fig.~\ref{fig:sw}(a), the lowest eigenvalues of $\Lambda_{\bm{k}}$ are found to be twofold degenerate at $\pm K$ valleys of HBZ. The classical ground state
	\begin{equation}
		\Psi\left(\bm{r}\right)=\frac{1}{\sqrt{2}}
		\left[e^{+i\varphi/2}\psi_{+K}\left(\bm{r}\right)
		+e^{-i\varphi/2}\psi_{-K}\left(\bm{r}\right)\right]
		\label{eq:cgs}
	\end{equation}
	can be constructed by the time-reversal symmetric intervalley coherent superposition of Bloch wave functions $\psi_{\pm K}$ with the lowest eigenvalues at $\pm K$ valleys on the equator of Bloch sphere, as shown in Fig.~\ref{fig:sw}(b). The pseudospins are then rotated about $y$ axis to the local frame with $z$ axis being the ordered direction
	\begin{equation}
		\left[
		\begin{matrix}
			T_z \\
			T_x  
		\end{matrix}
		\right]
		=
		\left[
		\begin{matrix}
			\cos\phi & \sin\phi \\
			-\sin\phi  & \cos\phi
		\end{matrix}
		\right]
		\left[
		\begin{matrix}
			\tau_z \\
			\tau_x  
		\end{matrix}
		\right].
		\label{eq:local}
	\end{equation}
	Correspondingly, the polar angle $\phi\left(\bm{r}\right)=-K\cdot\bm{r}-\varphi/2$ and $K\cdot\bm{r}+\varphi/2+\pi$ in $zx$ plane for sublattice $\mathcal{A}$ and $\mathcal{B}$, respectively. Note that the $\varphi$-continuous symmetry restricted to the classical ground state is emergent. The orbital pseudospin of the classical ground state evolves in the $zx$ plane without any energy cost, which makes the system particularly susceptible to quantum fluctuations. Following Holstein-Primakoff spin wave theory~\cite{Holstein1940}, the zero-point energy, arising from the correction of quantum fluctuations, is further studied as a function of coherent phase $\varphi$ in Eq.~(\ref{eq:cgs}). To the leading order, the zero-point energy takes the form
	\begin{equation}
		E_\text{ZP}(\varphi)=\frac{1}{N}\sum_{i\bm{k}}E_{i\bm{k}}\left(\varphi\right)-\frac{9}{2}J,
		\label{eq:zp}
	\end{equation}
	where $N$ is the number of unit cells, $E_{i\bm{k}}$ is $i$-th band of orbital excitations. The numerical evaluation of zero-point energy $E_\text{ZP}\left(\varphi\right)$ is shown in Fig.~\ref{fig:sw}(c). The orbital fluctuation lifts the continuous degeneracy, and selects a unique ground state at the coherent phase $\varphi=2\pi/3$ through the order-by-disorder mechanism in frustrated spin systems~\cite{Villain1977,Villain1980,Shender1982,Henley1989,Lacroix2011,Diep2012,Green2018}, recovering the configuration of orbital pseudospins predicted by the Hartree-Fock results.  
	For the selected ground state, the orbiton Hamiltonian is expressed as follow 	
	\begin{equation}
		H_\text{LOW}=\underset{\langle ij \rangle\in\varhexagon}{\sum}
		\frac{3}{16}J 
		(a_i^\dagger a_j + a_i^\dagger a_j^\dagger + \text{h.c.})
		+\sum_i \frac{3}{4}J a_i^\dagger a_i, 
		\label{eq:LSW}
	\end{equation}
	\begin{equation}
		H_\text{LOW}^\prime=
		\frac{J^\prime}{8} \underset{\langle ij \rangle\in\varhexagon}{\sum}
		(3a_i^\dagger a_j - a_i^\dagger a_j^\dagger)
		-\frac{J^\prime}{2}  \underset{\langle ij \rangle\notin\varhexagon}{\sum}
		a_i^\dagger a_j 
	 	+ \text{h.c.} 
		\label{eq:LSWP}
	\end{equation}
	by separating $J$ and $J^\prime$ superexchanges in Eq.~(\ref{eq:orbex}). Clearly, the dominate $J$ terms in Eq.~(\ref{eq:LSW}) highlight the Kekul\'e texture arising from the intervalley coherence.

	To summarize, the spinless fermions in the $p$-band honeycomb lattice undergo a sequence of DSM-QAHI-IVCI phase transitions in response to the Hubbard interaction. In particular, we have predicted a universal IVCI state that straddles the limits of Slater and Mott insulating phases.
	Our study illustrates the essential ingredients of design principles to guide the search of emergent Kekul\'e distortions in correlated multi-orbital systems. Ultracold atoms may offer an experimental platform with orbital degree of freedom in optical lattices, which is already demonstrated for bosons~\cite{Mueller2007,Wirth2010,SoltanPanahi2011,Oelschlaeger2013,Kock2015,Jin2021,Vargas2021,Wang2021,Wang2023} and is promising for fermions~\cite{Koehl2005} with related fast developing techniques~\cite{Mamaev2021,Venu2023}. One further direction is the generalization of our study to spin-1/2 fermions, which is relevant for solid-state materials. Our findings open new directions for the search of Kekul\'e orbitons in strongly correlated systems. 
	 
	We thank C.J. Wu, W.V. Liu and E.H. Zhao for helpful discussions. 
	This work is supported by National Natural Science Foundation of China under Grants No. 12174345, and Zhejiang Provincial Natural Science Foundation of China under Grant No. LZ22A040002. 
	

\begin{thebibliography}{70}%
		\makeatletter
		\providecommand \@ifxundefined [1]{%
			\@ifx{#1\undefined}
		}%
		\providecommand \@ifnum [1]{%
			\ifnum #1\expandafter \@firstoftwo
			\else \expandafter \@secondoftwo
			\fi
		}%
		\providecommand \@ifx [1]{%
			\ifx #1\expandafter \@firstoftwo
			\else \expandafter \@secondoftwo
			\fi
		}%
		\providecommand \natexlab [1]{#1}%
		\providecommand \enquote  [1]{``#1''}%
		\providecommand \bibnamefont  [1]{#1}%
		\providecommand \bibfnamefont [1]{#1}%
		\providecommand \citenamefont [1]{#1}%
		\providecommand \href@noop [0]{\@secondoftwo}%
		\providecommand \href [0]{\begingroup \@sanitize@url \@href}%
		\providecommand \@href[1]{\@@startlink{#1}\@@href}%
		\providecommand \@@href[1]{\endgroup#1\@@endlink}%
		\providecommand \@sanitize@url [0]{\catcode `\\12\catcode `\$12\catcode
			`\&12\catcode `\#12\catcode `\^12\catcode `\_12\catcode `\%12\relax}%
		\providecommand \@@startlink[1]{}%
		\providecommand \@@endlink[0]{}%
		\providecommand \url  [0]{\begingroup\@sanitize@url \@url }%
		\providecommand \@url [1]{\endgroup\@href {#1}{\urlprefix }}%
		\providecommand \urlprefix  [0]{URL }%
		\providecommand \Eprint [0]{\href }%
		\providecommand \doibase [0]{https://doi.org/}%
		\providecommand \selectlanguage [0]{\@gobble}%
		\providecommand \bibinfo  [0]{\@secondoftwo}%
		\providecommand \bibfield  [0]{\@secondoftwo}%
		\providecommand \translation [1]{[#1]}%
		\providecommand \BibitemOpen [0]{}%
		\providecommand \bibitemStop [0]{}%
		\providecommand \bibitemNoStop [0]{.\EOS\space}%
		\providecommand \EOS [0]{\spacefactor3000\relax}%
		\providecommand \BibitemShut  [1]{\csname bibitem#1\endcsname}%
		\let\auto@bib@innerbib\@empty
		\bibitem [{\citenamefont {Li}\ \emph {et~al.}(2019)\citenamefont {Li},
			\citenamefont {Zhang}, \citenamefont {Yin},\ and\ \citenamefont
			{He}}]{Li2019}%
		\BibitemOpen
		\bibfield  {author} {\bibinfo {author} {\bibfnamefont {S.-Y.}\ \bibnamefont
				{Li}}, \bibinfo {author} {\bibfnamefont {Y.}~\bibnamefont {Zhang}}, \bibinfo
			{author} {\bibfnamefont {L.-J.}\ \bibnamefont {Yin}},\ and\ \bibinfo {author}
			{\bibfnamefont {L.}~\bibnamefont {He}},\ }\href
		{https://doi.org/10.1103/PhysRevB.100.085437} {\bibfield  {journal} {\bibinfo
				{journal} {Phys. Rev. B}\ }\textbf {\bibinfo {volume} {100}},\ \bibinfo
			{pages} {085437} (\bibinfo {year} {2019})}\BibitemShut {NoStop}%
		\bibitem [{\citenamefont {Liu}\ \emph {et~al.}(2022)\citenamefont {Liu},
			\citenamefont {Farahi}, \citenamefont {Chiu}, \citenamefont {Papic},
			\citenamefont {Watanabe}, \citenamefont {Taniguchi}, \citenamefont
			{Zaletel},\ and\ \citenamefont {Yazdani}}]{Liu2022a}%
		\BibitemOpen
		\bibfield  {author} {\bibinfo {author} {\bibfnamefont {X.}~\bibnamefont
				{Liu}}, \bibinfo {author} {\bibfnamefont {G.}~\bibnamefont {Farahi}},
			\bibinfo {author} {\bibfnamefont {C.-L.}\ \bibnamefont {Chiu}}, \bibinfo
			{author} {\bibfnamefont {Z.}~\bibnamefont {Papic}}, \bibinfo {author}
			{\bibfnamefont {K.}~\bibnamefont {Watanabe}}, \bibinfo {author}
			{\bibfnamefont {T.}~\bibnamefont {Taniguchi}}, \bibinfo {author}
			{\bibfnamefont {M.~P.}\ \bibnamefont {Zaletel}},\ and\ \bibinfo {author}
			{\bibfnamefont {A.}~\bibnamefont {Yazdani}},\ }\href
		{https://doi.org/10.1126/science.abm3770} {\bibfield  {journal} {\bibinfo
				{journal} {Science}\ }\textbf {\bibinfo {volume} {375}},\ \bibinfo {pages}
			{321} (\bibinfo {year} {2022})}\BibitemShut {NoStop}%
		\bibitem [{\citenamefont {Coissard}\ \emph {et~al.}(2022)\citenamefont
			{Coissard}, \citenamefont {Wander}, \citenamefont {Vignaud}, \citenamefont
			{Grushin}, \citenamefont {Repellin}, \citenamefont {Watanabe}, \citenamefont
			{Taniguchi}, \citenamefont {Gay}, \citenamefont {Winkelmann}, \citenamefont
			{Courtois}, \citenamefont {Sellier},\ and\ \citenamefont
			{Sacépé}}]{Coissard2022}%
		\BibitemOpen
		\bibfield  {author} {\bibinfo {author} {\bibfnamefont {A.}~\bibnamefont
				{Coissard}}, \bibinfo {author} {\bibfnamefont {D.}~\bibnamefont {Wander}},
			\bibinfo {author} {\bibfnamefont {H.}~\bibnamefont {Vignaud}}, \bibinfo
			{author} {\bibfnamefont {A.~G.}\ \bibnamefont {Grushin}}, \bibinfo {author}
			{\bibfnamefont {C.}~\bibnamefont {Repellin}}, \bibinfo {author}
			{\bibfnamefont {K.}~\bibnamefont {Watanabe}}, \bibinfo {author}
			{\bibfnamefont {T.}~\bibnamefont {Taniguchi}}, \bibinfo {author}
			{\bibfnamefont {F.}~\bibnamefont {Gay}}, \bibinfo {author} {\bibfnamefont
				{C.~B.}\ \bibnamefont {Winkelmann}}, \bibinfo {author} {\bibfnamefont
				{H.}~\bibnamefont {Courtois}}, \bibinfo {author} {\bibfnamefont
				{H.}~\bibnamefont {Sellier}},\ and\ \bibinfo {author} {\bibfnamefont
				{B.}~\bibnamefont {Sacépé}},\ }\href
		{https://doi.org/10.1038/s41586-022-04513-7} {\bibfield  {journal} {\bibinfo
				{journal} {Nature}\ }\textbf {\bibinfo {volume} {605}},\ \bibinfo {pages}
			{51} (\bibinfo {year} {2022})}\BibitemShut {NoStop}%
		\bibitem [{\citenamefont {Nuckolls}\ \emph {et~al.}(2023)\citenamefont
			{Nuckolls}, \citenamefont {Lee}, \citenamefont {Oh}, \citenamefont {Wong},
			\citenamefont {Soejima}, \citenamefont {Hong}, \citenamefont {Călugăru},
			\citenamefont {Herzog-Arbeitman}, \citenamefont {Bernevig}, \citenamefont
			{Watanabe}, \citenamefont {Taniguchi}, \citenamefont {Regnault},
			\citenamefont {Zaletel},\ and\ \citenamefont {Yazdani}}]{Nuckolls2023}%
		\BibitemOpen
		\bibfield  {author} {\bibinfo {author} {\bibfnamefont {K.~P.}\ \bibnamefont
				{Nuckolls}}, \bibinfo {author} {\bibfnamefont {R.~L.}\ \bibnamefont {Lee}},
			\bibinfo {author} {\bibfnamefont {M.}~\bibnamefont {Oh}}, \bibinfo {author}
			{\bibfnamefont {D.}~\bibnamefont {Wong}}, \bibinfo {author} {\bibfnamefont
				{T.}~\bibnamefont {Soejima}}, \bibinfo {author} {\bibfnamefont {J.~P.}\
				\bibnamefont {Hong}}, \bibinfo {author} {\bibfnamefont {D.}~\bibnamefont
				{Călugăru}}, \bibinfo {author} {\bibfnamefont {J.}~\bibnamefont
				{Herzog-Arbeitman}}, \bibinfo {author} {\bibfnamefont {B.~A.}\ \bibnamefont
				{Bernevig}}, \bibinfo {author} {\bibfnamefont {K.}~\bibnamefont {Watanabe}},
			\bibinfo {author} {\bibfnamefont {T.}~\bibnamefont {Taniguchi}}, \bibinfo
			{author} {\bibfnamefont {N.}~\bibnamefont {Regnault}}, \bibinfo {author}
			{\bibfnamefont {M.~P.}\ \bibnamefont {Zaletel}},\ and\ \bibinfo {author}
			{\bibfnamefont {A.}~\bibnamefont {Yazdani}},\ }\href
		{https://doi.org/10.1038/s41586-023-06226-x} {\bibfield  {journal} {\bibinfo
				{journal} {Nature}\ }\textbf {\bibinfo {volume} {620}},\ \bibinfo {pages}
			{525} (\bibinfo {year} {2023})}\BibitemShut {NoStop}%
		\bibitem [{\citenamefont {Kim}\ \emph {et~al.}(2023)\citenamefont {Kim},
			\citenamefont {Choi}, \citenamefont {Lantagne-Hurtubise}, \citenamefont
			{Lewandowski}, \citenamefont {Thomson}, \citenamefont {Kong}, \citenamefont
			{Zhou}, \citenamefont {Baum}, \citenamefont {Zhang}, \citenamefont {Holleis},
			\citenamefont {Watanabe}, \citenamefont {Taniguchi}, \citenamefont {Young},
			\citenamefont {Alicea},\ and\ \citenamefont {Nadj-Perge}}]{Kim2023}%
		\BibitemOpen
		\bibfield  {author} {\bibinfo {author} {\bibfnamefont {H.}~\bibnamefont
				{Kim}}, \bibinfo {author} {\bibfnamefont {Y.}~\bibnamefont {Choi}}, \bibinfo
			{author} {\bibfnamefont {E.}~\bibnamefont {Lantagne-Hurtubise}}, \bibinfo
			{author} {\bibfnamefont {C.}~\bibnamefont {Lewandowski}}, \bibinfo {author}
			{\bibfnamefont {A.}~\bibnamefont {Thomson}}, \bibinfo {author} {\bibfnamefont
				{L.}~\bibnamefont {Kong}}, \bibinfo {author} {\bibfnamefont {H.}~\bibnamefont
				{Zhou}}, \bibinfo {author} {\bibfnamefont {E.}~\bibnamefont {Baum}}, \bibinfo
			{author} {\bibfnamefont {Y.}~\bibnamefont {Zhang}}, \bibinfo {author}
			{\bibfnamefont {L.}~\bibnamefont {Holleis}}, \bibinfo {author} {\bibfnamefont
				{K.}~\bibnamefont {Watanabe}}, \bibinfo {author} {\bibfnamefont
				{T.}~\bibnamefont {Taniguchi}}, \bibinfo {author} {\bibfnamefont {A.~F.}\
				\bibnamefont {Young}}, \bibinfo {author} {\bibfnamefont {J.}~\bibnamefont
				{Alicea}},\ and\ \bibinfo {author} {\bibfnamefont {S.}~\bibnamefont
				{Nadj-Perge}},\ }\href {https://doi.org/10.1038/s41586-023-06663-8}
		{\bibfield  {journal} {\bibinfo  {journal} {Nature}\ }\textbf {\bibinfo
				{volume} {623}},\ \bibinfo {pages} {942} (\bibinfo {year}
			{2023})}\BibitemShut {NoStop}%
		\bibitem [{\citenamefont {Bultinck}\ \emph {et~al.}(2020)\citenamefont
			{Bultinck}, \citenamefont {Khalaf}, \citenamefont {Liu}, \citenamefont
			{Chatterjee}, \citenamefont {Vishwanath},\ and\ \citenamefont
			{Zaletel}}]{Bultinck2020}%
		\BibitemOpen
		\bibfield  {author} {\bibinfo {author} {\bibfnamefont {N.}~\bibnamefont
				{Bultinck}}, \bibinfo {author} {\bibfnamefont {E.}~\bibnamefont {Khalaf}},
			\bibinfo {author} {\bibfnamefont {S.}~\bibnamefont {Liu}}, \bibinfo {author}
			{\bibfnamefont {S.}~\bibnamefont {Chatterjee}}, \bibinfo {author}
			{\bibfnamefont {A.}~\bibnamefont {Vishwanath}},\ and\ \bibinfo {author}
			{\bibfnamefont {M.~P.}\ \bibnamefont {Zaletel}},\ }\href
		{https://doi.org/10.1103/PhysRevX.10.031034} {\bibfield  {journal} {\bibinfo
				{journal} {Phys. Rev. X}\ }\textbf {\bibinfo {volume} {10}},\ \bibinfo
			{pages} {031034} (\bibinfo {year} {2020})}\BibitemShut {NoStop}%
		\bibitem [{\citenamefont {Atteia}\ \emph {et~al.}(2021)\citenamefont {Atteia},
			\citenamefont {Lian},\ and\ \citenamefont {Goerbig}}]{Atteia2021}%
		\BibitemOpen
		\bibfield  {author} {\bibinfo {author} {\bibfnamefont {J.}~\bibnamefont
				{Atteia}}, \bibinfo {author} {\bibfnamefont {Y.}~\bibnamefont {Lian}},\ and\
			\bibinfo {author} {\bibfnamefont {M.~O.}\ \bibnamefont {Goerbig}},\ }\href
		{https://doi.org/10.1103/PhysRevB.103.035403} {\bibfield  {journal} {\bibinfo
				{journal} {Phys. Rev. B}\ }\textbf {\bibinfo {volume} {103}},\ \bibinfo
			{pages} {035403} (\bibinfo {year} {2021})}\BibitemShut {NoStop}%
		\bibitem [{\citenamefont {Kwan}\ \emph {et~al.}(2021)\citenamefont {Kwan},
			\citenamefont {Wagner}, \citenamefont {Soejima}, \citenamefont {Zaletel},
			\citenamefont {Simon}, \citenamefont {Parameswaran},\ and\ \citenamefont
			{Bultinck}}]{Kwan2021}%
		\BibitemOpen
		\bibfield  {author} {\bibinfo {author} {\bibfnamefont {Y.~H.}\ \bibnamefont
				{Kwan}}, \bibinfo {author} {\bibfnamefont {G.}~\bibnamefont {Wagner}},
			\bibinfo {author} {\bibfnamefont {T.}~\bibnamefont {Soejima}}, \bibinfo
			{author} {\bibfnamefont {M.~P.}\ \bibnamefont {Zaletel}}, \bibinfo {author}
			{\bibfnamefont {S.~H.}\ \bibnamefont {Simon}}, \bibinfo {author}
			{\bibfnamefont {S.~A.}\ \bibnamefont {Parameswaran}},\ and\ \bibinfo {author}
			{\bibfnamefont {N.}~\bibnamefont {Bultinck}},\ }\href
		{https://doi.org/10.1103/PhysRevX.11.041063} {\bibfield  {journal} {\bibinfo
				{journal} {Phys. Rev. X}\ }\textbf {\bibinfo {volume} {11}},\ \bibinfo
			{pages} {041063} (\bibinfo {year} {2021})}\BibitemShut {NoStop}%
		\bibitem [{\citenamefont {C\ifmmode \u{a}\else \u{a}\fi{}lug\ifmmode~\u{a}\else
				\u{a}\fi{}ru}\ \emph {et~al.}(2022)\citenamefont {C\ifmmode \u{a}\else
				\u{a}\fi{}lug\ifmmode~\u{a}\else \u{a}\fi{}ru}, \citenamefont {Regnault},
			\citenamefont {Oh}, \citenamefont {Nuckolls}, \citenamefont {Wong},
			\citenamefont {Lee}, \citenamefont {Yazdani}, \citenamefont {Vafek},\ and\
			\citenamefont {Bernevig}}]{Cifmmode2022}%
		\BibitemOpen
		\bibfield  {author} {\bibinfo {author} {\bibfnamefont {D.}~\bibnamefont
				{C\ifmmode \u{a}\else \u{a}\fi{}lug\ifmmode~\u{a}\else \u{a}\fi{}ru}},
			\bibinfo {author} {\bibfnamefont {N.}~\bibnamefont {Regnault}}, \bibinfo
			{author} {\bibfnamefont {M.}~\bibnamefont {Oh}}, \bibinfo {author}
			{\bibfnamefont {K.~P.}\ \bibnamefont {Nuckolls}}, \bibinfo {author}
			{\bibfnamefont {D.}~\bibnamefont {Wong}}, \bibinfo {author} {\bibfnamefont
				{R.~L.}\ \bibnamefont {Lee}}, \bibinfo {author} {\bibfnamefont
				{A.}~\bibnamefont {Yazdani}}, \bibinfo {author} {\bibfnamefont
				{O.}~\bibnamefont {Vafek}},\ and\ \bibinfo {author} {\bibfnamefont {B.~A.}\
				\bibnamefont {Bernevig}},\ }\href
		{https://doi.org/10.1103/PhysRevLett.129.117602} {\bibfield  {journal}
			{\bibinfo  {journal} {Phys. Rev. Lett.}\ }\textbf {\bibinfo {volume} {129}},\
			\bibinfo {pages} {117602} (\bibinfo {year} {2022})}\BibitemShut {NoStop}%
		\bibitem [{\citenamefont {Hong}\ \emph {et~al.}(2022)\citenamefont {Hong},
			\citenamefont {Soejima},\ and\ \citenamefont {Zaletel}}]{Hong2022}%
		\BibitemOpen
		\bibfield  {author} {\bibinfo {author} {\bibfnamefont {J.~P.}\ \bibnamefont
				{Hong}}, \bibinfo {author} {\bibfnamefont {T.}~\bibnamefont {Soejima}},\ and\
			\bibinfo {author} {\bibfnamefont {M.~P.}\ \bibnamefont {Zaletel}},\ }\href
		{https://doi.org/10.1103/PhysRevLett.129.147001} {\bibfield  {journal}
			{\bibinfo  {journal} {Phys. Rev. Lett.}\ }\textbf {\bibinfo {volume} {129}},\
			\bibinfo {pages} {147001} (\bibinfo {year} {2022})}\BibitemShut {NoStop}%
		\bibitem [{\citenamefont {Christos}\ \emph {et~al.}(2022)\citenamefont
			{Christos}, \citenamefont {Sachdev},\ and\ \citenamefont
			{Scheurer}}]{Christos2022}%
		\BibitemOpen
		\bibfield  {author} {\bibinfo {author} {\bibfnamefont {M.}~\bibnamefont
				{Christos}}, \bibinfo {author} {\bibfnamefont {S.}~\bibnamefont {Sachdev}},\
			and\ \bibinfo {author} {\bibfnamefont {M.~S.}\ \bibnamefont {Scheurer}},\
		}\href {https://doi.org/10.1103/PhysRevX.12.021018} {\bibfield  {journal}
			{\bibinfo  {journal} {Phys. Rev. X}\ }\textbf {\bibinfo {volume} {12}},\
			\bibinfo {pages} {021018} (\bibinfo {year} {2022})}\BibitemShut {NoStop}%
		\bibitem [{\citenamefont {Das}\ \emph {et~al.}(2022)\citenamefont {Das},
			\citenamefont {Kaul},\ and\ \citenamefont {Murthy}}]{Das2022}%
		\BibitemOpen
		\bibfield  {author} {\bibinfo {author} {\bibfnamefont {A.}~\bibnamefont
				{Das}}, \bibinfo {author} {\bibfnamefont {R.~K.}\ \bibnamefont {Kaul}},\ and\
			\bibinfo {author} {\bibfnamefont {G.}~\bibnamefont {Murthy}},\ }\href
		{https://doi.org/10.1103/PhysRevLett.128.106803} {\bibfield  {journal}
			{\bibinfo  {journal} {Phys. Rev. Lett.}\ }\textbf {\bibinfo {volume} {128}},\
			\bibinfo {pages} {106803} (\bibinfo {year} {2022})}\BibitemShut {NoStop}%
		\bibitem [{\citenamefont {You}\ and\ \citenamefont
			{Vishwanath}(2022)}]{You2022}%
		\BibitemOpen
		\bibfield  {author} {\bibinfo {author} {\bibfnamefont {Y.-Z.}\ \bibnamefont
				{You}}\ and\ \bibinfo {author} {\bibfnamefont {A.}~\bibnamefont
				{Vishwanath}},\ }\href {https://doi.org/10.1103/PhysRevB.105.134524}
		{\bibfield  {journal} {\bibinfo  {journal} {Phys. Rev. B}\ }\textbf {\bibinfo
				{volume} {105}},\ \bibinfo {pages} {134524} (\bibinfo {year}
			{2022})}\BibitemShut {NoStop}%
		\bibitem [{\citenamefont {S\'anchez~S\'anchez}\ \emph
			{et~al.}(2024)\citenamefont {S\'anchez~S\'anchez}, \citenamefont {D\'{\i}az},
			\citenamefont {Gonz\'alez},\ and\ \citenamefont {Stauber}}]{Sanchez2024}%
		\BibitemOpen
		\bibfield  {author} {\bibinfo {author} {\bibfnamefont {M.}~\bibnamefont
				{S\'anchez~S\'anchez}}, \bibinfo {author} {\bibfnamefont {I.}~\bibnamefont
				{D\'{\i}az}}, \bibinfo {author} {\bibfnamefont {J.}~\bibnamefont
				{Gonz\'alez}},\ and\ \bibinfo {author} {\bibfnamefont {T.}~\bibnamefont
				{Stauber}},\ }\href {https://doi.org/10.1103/PhysRevLett.133.266603}
		{\bibfield  {journal} {\bibinfo  {journal} {Phys. Rev. Lett.}\ }\textbf
			{\bibinfo {volume} {133}},\ \bibinfo {pages} {266603} (\bibinfo {year}
			{2024})}\BibitemShut {NoStop}%
		\bibitem [{\citenamefont {Ajiki}\ and\ \citenamefont {Ando}(1995)}]{Ajiki1995}%
		\BibitemOpen
		\bibfield  {author} {\bibinfo {author} {\bibfnamefont {H.}~\bibnamefont
				{Ajiki}}\ and\ \bibinfo {author} {\bibfnamefont {T.}~\bibnamefont {Ando}},\
		}\href {https://doi.org/10.1143/jpsj.64.260} {\bibfield  {journal} {\bibinfo
				{journal} {Journal of the Physical Society of Japan}\ }\textbf {\bibinfo
				{volume} {64}},\ \bibinfo {pages} {260} (\bibinfo {year} {1995})}\BibitemShut
		{NoStop}%
		\bibitem [{\citenamefont {Chamon}(2000)}]{Chamon2000}%
		\BibitemOpen
		\bibfield  {author} {\bibinfo {author} {\bibfnamefont {C.}~\bibnamefont
				{Chamon}},\ }\href {https://doi.org/10.1103/PhysRevB.62.2806} {\bibfield
			{journal} {\bibinfo  {journal} {Phys. Rev. B}\ }\textbf {\bibinfo {volume}
				{62}},\ \bibinfo {pages} {2806} (\bibinfo {year} {2000})}\BibitemShut
		{NoStop}%
		\bibitem [{\citenamefont {Hou}\ \emph {et~al.}(2007)\citenamefont {Hou},
			\citenamefont {Chamon},\ and\ \citenamefont {Mudry}}]{Hou2007}%
		\BibitemOpen
		\bibfield  {author} {\bibinfo {author} {\bibfnamefont {C.-Y.}\ \bibnamefont
				{Hou}}, \bibinfo {author} {\bibfnamefont {C.}~\bibnamefont {Chamon}},\ and\
			\bibinfo {author} {\bibfnamefont {C.}~\bibnamefont {Mudry}},\ }\href
		{https://doi.org/10.1103/PhysRevLett.98.186809} {\bibfield  {journal}
			{\bibinfo  {journal} {Phys. Rev. Lett.}\ }\textbf {\bibinfo {volume} {98}},\
			\bibinfo {pages} {186809} (\bibinfo {year} {2007})}\BibitemShut {NoStop}%
		\bibitem [{\citenamefont {Frank}\ and\ \citenamefont {Lieb}(2011)}]{Frank2011}%
		\BibitemOpen
		\bibfield  {author} {\bibinfo {author} {\bibfnamefont {R.~L.}\ \bibnamefont
				{Frank}}\ and\ \bibinfo {author} {\bibfnamefont {E.~H.}\ \bibnamefont
				{Lieb}},\ }\href {https://doi.org/10.1103/PhysRevLett.107.066801} {\bibfield
			{journal} {\bibinfo  {journal} {Phys. Rev. Lett.}\ }\textbf {\bibinfo
				{volume} {107}},\ \bibinfo {pages} {066801} (\bibinfo {year}
			{2011})}\BibitemShut {NoStop}%
		\bibitem [{\citenamefont {Wu}\ and\ \citenamefont {Hu}(2015)}]{Wu2015}%
		\BibitemOpen
		\bibfield  {author} {\bibinfo {author} {\bibfnamefont {L.-H.}\ \bibnamefont
				{Wu}}\ and\ \bibinfo {author} {\bibfnamefont {X.}~\bibnamefont {Hu}},\ }\href
		{https://doi.org/10.1103/PhysRevLett.114.223901} {\bibfield  {journal}
			{\bibinfo  {journal} {Phys. Rev. Lett.}\ }\textbf {\bibinfo {volume} {114}},\
			\bibinfo {pages} {223901} (\bibinfo {year} {2015})}\BibitemShut {NoStop}%
		\bibitem [{\citenamefont {Gutiérrez}\ \emph {et~al.}(2016)\citenamefont
			{Gutiérrez}, \citenamefont {Kim}, \citenamefont {Brown}, \citenamefont
			{Schiros}, \citenamefont {Nordlund}, \citenamefont {Lochocki}, \citenamefont
			{Shen}, \citenamefont {Park},\ and\ \citenamefont
			{Pasupathy}}]{Gutierrez2016}%
		\BibitemOpen
		\bibfield  {author} {\bibinfo {author} {\bibfnamefont {C.}~\bibnamefont
				{Gutiérrez}}, \bibinfo {author} {\bibfnamefont {C.-J.}\ \bibnamefont {Kim}},
			\bibinfo {author} {\bibfnamefont {L.}~\bibnamefont {Brown}}, \bibinfo
			{author} {\bibfnamefont {T.}~\bibnamefont {Schiros}}, \bibinfo {author}
			{\bibfnamefont {D.}~\bibnamefont {Nordlund}}, \bibinfo {author}
			{\bibfnamefont {E.}~\bibnamefont {Lochocki}}, \bibinfo {author}
			{\bibfnamefont {K.~M.}\ \bibnamefont {Shen}}, \bibinfo {author}
			{\bibfnamefont {J.}~\bibnamefont {Park}},\ and\ \bibinfo {author}
			{\bibfnamefont {A.~N.}\ \bibnamefont {Pasupathy}},\ }\href
		{https://doi.org/10.1038/nphys3776} {\bibfield  {journal} {\bibinfo
				{journal} {Nature Physics}\ }\textbf {\bibinfo {volume} {12}},\ \bibinfo
			{pages} {950} (\bibinfo {year} {2016})}\BibitemShut {NoStop}%
		\bibitem [{\citenamefont {Liu}\ \emph {et~al.}(2017)\citenamefont {Liu},
			\citenamefont {Lian}, \citenamefont {Li}, \citenamefont {Xu},\ and\
			\citenamefont {Duan}}]{Liu2017a}%
		\BibitemOpen
		\bibfield  {author} {\bibinfo {author} {\bibfnamefont {Y.}~\bibnamefont
				{Liu}}, \bibinfo {author} {\bibfnamefont {C.-S.}\ \bibnamefont {Lian}},
			\bibinfo {author} {\bibfnamefont {Y.}~\bibnamefont {Li}}, \bibinfo {author}
			{\bibfnamefont {Y.}~\bibnamefont {Xu}},\ and\ \bibinfo {author}
			{\bibfnamefont {W.}~\bibnamefont {Duan}},\ }\href
		{https://doi.org/10.1103/PhysRevLett.119.255901} {\bibfield  {journal}
			{\bibinfo  {journal} {Phys. Rev. Lett.}\ }\textbf {\bibinfo {volume} {119}},\
			\bibinfo {pages} {255901} (\bibinfo {year} {2017})}\BibitemShut {NoStop}%
		\bibitem [{\citenamefont {Gao}\ \emph {et~al.}(2019)\citenamefont {Gao},
			\citenamefont {Torrent}, \citenamefont {Cervera}, \citenamefont {San-Jose},
			\citenamefont {S\'anchez-Dehesa},\ and\ \citenamefont
			{Christensen}}]{Gao2019}%
		\BibitemOpen
		\bibfield  {author} {\bibinfo {author} {\bibfnamefont {P.}~\bibnamefont
				{Gao}}, \bibinfo {author} {\bibfnamefont {D.}~\bibnamefont {Torrent}},
			\bibinfo {author} {\bibfnamefont {F.}~\bibnamefont {Cervera}}, \bibinfo
			{author} {\bibfnamefont {P.}~\bibnamefont {San-Jose}}, \bibinfo {author}
			{\bibfnamefont {J.}~\bibnamefont {S\'anchez-Dehesa}},\ and\ \bibinfo {author}
			{\bibfnamefont {J.}~\bibnamefont {Christensen}},\ }\href
		{https://doi.org/10.1103/PhysRevLett.123.196601} {\bibfield  {journal}
			{\bibinfo  {journal} {Phys. Rev. Lett.}\ }\textbf {\bibinfo {volume} {123}},\
			\bibinfo {pages} {196601} (\bibinfo {year} {2019})}\BibitemShut {NoStop}%
		\bibitem [{\citenamefont {Menssen}\ \emph {et~al.}(2020)\citenamefont
			{Menssen}, \citenamefont {Guan}, \citenamefont {Felce}, \citenamefont
			{Booth},\ and\ \citenamefont {Walmsley}}]{Menssen2020}%
		\BibitemOpen
		\bibfield  {author} {\bibinfo {author} {\bibfnamefont {A.~J.}\ \bibnamefont
				{Menssen}}, \bibinfo {author} {\bibfnamefont {J.}~\bibnamefont {Guan}},
			\bibinfo {author} {\bibfnamefont {D.}~\bibnamefont {Felce}}, \bibinfo
			{author} {\bibfnamefont {M.~J.}\ \bibnamefont {Booth}},\ and\ \bibinfo
			{author} {\bibfnamefont {I.~A.}\ \bibnamefont {Walmsley}},\ }\href
		{https://doi.org/10.1103/PhysRevLett.125.117401} {\bibfield  {journal}
			{\bibinfo  {journal} {Phys. Rev. Lett.}\ }\textbf {\bibinfo {volume} {125}},\
			\bibinfo {pages} {117401} (\bibinfo {year} {2020})}\BibitemShut {NoStop}%
		\bibitem [{\citenamefont {Gao}\ \emph {et~al.}(2020)\citenamefont {Gao},
			\citenamefont {Yang}, \citenamefont {Lin}, \citenamefont {Zhang},
			\citenamefont {Li}, \citenamefont {Bo}, \citenamefont {Wang},\ and\
			\citenamefont {Lu}}]{Gao2020}%
		\BibitemOpen
		\bibfield  {author} {\bibinfo {author} {\bibfnamefont {X.}~\bibnamefont
				{Gao}}, \bibinfo {author} {\bibfnamefont {L.}~\bibnamefont {Yang}}, \bibinfo
			{author} {\bibfnamefont {H.}~\bibnamefont {Lin}}, \bibinfo {author}
			{\bibfnamefont {L.}~\bibnamefont {Zhang}}, \bibinfo {author} {\bibfnamefont
				{J.}~\bibnamefont {Li}}, \bibinfo {author} {\bibfnamefont {F.}~\bibnamefont
				{Bo}}, \bibinfo {author} {\bibfnamefont {Z.}~\bibnamefont {Wang}},\ and\
			\bibinfo {author} {\bibfnamefont {L.}~\bibnamefont {Lu}},\ }\href
		{https://doi.org/10.1038/s41565-020-0773-7} {\bibfield  {journal} {\bibinfo
				{journal} {Nature Nanotechnology}\ }\textbf {\bibinfo {volume} {15}},\
			\bibinfo {pages} {1012} (\bibinfo {year} {2020})}\BibitemShut {NoStop}%
		\bibitem [{\citenamefont {Bao}\ \emph {et~al.}(2021)\citenamefont {Bao},
			\citenamefont {Zhang}, \citenamefont {Zhang}, \citenamefont {Wu},
			\citenamefont {Luo}, \citenamefont {Zhou}, \citenamefont {Li}, \citenamefont
			{Hou}, \citenamefont {Yao}, \citenamefont {Liu}, \citenamefont {Yu},
			\citenamefont {Li}, \citenamefont {Duan}, \citenamefont {Yao}, \citenamefont
			{Wang},\ and\ \citenamefont {Zhou}}]{Bao2021}%
		\BibitemOpen
		\bibfield  {author} {\bibinfo {author} {\bibfnamefont {C.}~\bibnamefont
				{Bao}}, \bibinfo {author} {\bibfnamefont {H.}~\bibnamefont {Zhang}}, \bibinfo
			{author} {\bibfnamefont {T.}~\bibnamefont {Zhang}}, \bibinfo {author}
			{\bibfnamefont {X.}~\bibnamefont {Wu}}, \bibinfo {author} {\bibfnamefont
				{L.}~\bibnamefont {Luo}}, \bibinfo {author} {\bibfnamefont {S.}~\bibnamefont
				{Zhou}}, \bibinfo {author} {\bibfnamefont {Q.}~\bibnamefont {Li}}, \bibinfo
			{author} {\bibfnamefont {Y.}~\bibnamefont {Hou}}, \bibinfo {author}
			{\bibfnamefont {W.}~\bibnamefont {Yao}}, \bibinfo {author} {\bibfnamefont
				{L.}~\bibnamefont {Liu}}, \bibinfo {author} {\bibfnamefont {P.}~\bibnamefont
				{Yu}}, \bibinfo {author} {\bibfnamefont {J.}~\bibnamefont {Li}}, \bibinfo
			{author} {\bibfnamefont {W.}~\bibnamefont {Duan}}, \bibinfo {author}
			{\bibfnamefont {H.}~\bibnamefont {Yao}}, \bibinfo {author} {\bibfnamefont
				{Y.}~\bibnamefont {Wang}},\ and\ \bibinfo {author} {\bibfnamefont
				{S.}~\bibnamefont {Zhou}},\ }\href
		{https://doi.org/10.1103/PhysRevLett.126.206804} {\bibfield  {journal}
			{\bibinfo  {journal} {Phys. Rev. Lett.}\ }\textbf {\bibinfo {volume} {126}},\
			\bibinfo {pages} {206804} (\bibinfo {year} {2021})}\BibitemShut {NoStop}%
		\bibitem [{\citenamefont {Ma}\ \emph {et~al.}(2021)\citenamefont {Ma},
			\citenamefont {Xi}, \citenamefont {Li},\ and\ \citenamefont {Sun}}]{Ma2021}%
		\BibitemOpen
		\bibfield  {author} {\bibinfo {author} {\bibfnamefont {J.}~\bibnamefont
				{Ma}}, \bibinfo {author} {\bibfnamefont {X.}~\bibnamefont {Xi}}, \bibinfo
			{author} {\bibfnamefont {Y.}~\bibnamefont {Li}},\ and\ \bibinfo {author}
			{\bibfnamefont {X.}~\bibnamefont {Sun}},\ }\href
		{https://doi.org/10.1038/s41565-021-00868-6} {\bibfield  {journal} {\bibinfo
				{journal} {Nature Nanotechnology}\ }\textbf {\bibinfo {volume} {16}},\
			\bibinfo {pages} {576} (\bibinfo {year} {2021})}\BibitemShut {NoStop}%
		\bibitem [{\citenamefont {Guan}\ \emph {et~al.}(2024)\citenamefont {Guan},
			\citenamefont {Dutreix}, \citenamefont {González-Herrero}, \citenamefont
			{Ugeda}, \citenamefont {Brihuega}, \citenamefont {Katsnelson}, \citenamefont
			{Yazyev},\ and\ \citenamefont {Renard}}]{Guan2024}%
		\BibitemOpen
		\bibfield  {author} {\bibinfo {author} {\bibfnamefont {Y.}~\bibnamefont
				{Guan}}, \bibinfo {author} {\bibfnamefont {C.}~\bibnamefont {Dutreix}},
			\bibinfo {author} {\bibfnamefont {H.}~\bibnamefont {González-Herrero}},
			\bibinfo {author} {\bibfnamefont {M.~M.}\ \bibnamefont {Ugeda}}, \bibinfo
			{author} {\bibfnamefont {I.}~\bibnamefont {Brihuega}}, \bibinfo {author}
			{\bibfnamefont {M.~I.}\ \bibnamefont {Katsnelson}}, \bibinfo {author}
			{\bibfnamefont {O.~V.}\ \bibnamefont {Yazyev}},\ and\ \bibinfo {author}
			{\bibfnamefont {V.~T.}\ \bibnamefont {Renard}},\ }\bibfield  {journal}
		{\bibinfo  {journal} {Nature Communications}\ }\textbf {\bibinfo {volume}
			{15}},\ \href {https://doi.org/10.1038/s41467-024-47267-8}
		{10.1038/s41467-024-47267-8} (\bibinfo {year} {2024})\BibitemShut {NoStop}%
		\bibitem [{\citenamefont {Kugel’}\ and\ \citenamefont
			{Khomskiĭ}(1982)}]{Kugel1982}%
		\BibitemOpen
		\bibfield  {author} {\bibinfo {author} {\bibfnamefont {K.~I.}\ \bibnamefont
				{Kugel’}}\ and\ \bibinfo {author} {\bibfnamefont {D.~I.}\ \bibnamefont
				{Khomskiĭ}},\ }\href {https://doi.org/10.1070/pu1982v025n04abeh004537}
		{\bibfield  {journal} {\bibinfo  {journal} {Soviet Physics Uspekhi}\ }\textbf
			{\bibinfo {volume} {25}},\ \bibinfo {pages} {231} (\bibinfo {year}
			{1982})}\BibitemShut {NoStop}%
		\bibitem [{\citenamefont {van~den Brink}\ \emph {et~al.}(1999)\citenamefont
			{van~den Brink}, \citenamefont {Horsch}, \citenamefont {Mack},\ and\
			\citenamefont {Ole\ifmmode~\acute{s}\else \'{s}\fi{}}}]{Brink1999}%
		\BibitemOpen
		\bibfield  {author} {\bibinfo {author} {\bibfnamefont {J.}~\bibnamefont
				{van~den Brink}}, \bibinfo {author} {\bibfnamefont {P.}~\bibnamefont
				{Horsch}}, \bibinfo {author} {\bibfnamefont {F.}~\bibnamefont {Mack}},\ and\
			\bibinfo {author} {\bibfnamefont {A.~M.}\ \bibnamefont
				{Ole\ifmmode~\acute{s}\else \'{s}\fi{}}},\ }\href
		{https://doi.org/10.1103/PhysRevB.59.6795} {\bibfield  {journal} {\bibinfo
				{journal} {Phys. Rev. B}\ }\textbf {\bibinfo {volume} {59}},\ \bibinfo
			{pages} {6795} (\bibinfo {year} {1999})}\BibitemShut {NoStop}%
		\bibitem [{\citenamefont {Brink}(2004)}]{Brink2004}%
		\BibitemOpen
		\bibfield  {author} {\bibinfo {author} {\bibfnamefont {J.~v.~d.}\
				\bibnamefont {Brink}},\ }\href {https://doi.org/10.1088/1367-2630/6/1/201}
		{\bibfield  {journal} {\bibinfo  {journal} {New Journal of Physics}\ }\textbf
			{\bibinfo {volume} {6}},\ \bibinfo {pages} {201} (\bibinfo {year}
			{2004})}\BibitemShut {NoStop}%
		\bibitem [{\citenamefont {Nussinov}\ and\ \citenamefont {van~den
				Brink}(2015)}]{Nussinov2015}%
		\BibitemOpen
		\bibfield  {author} {\bibinfo {author} {\bibfnamefont {Z.}~\bibnamefont
				{Nussinov}}\ and\ \bibinfo {author} {\bibfnamefont {J.}~\bibnamefont {van~den
					Brink}},\ }\href {https://doi.org/10.1103/RevModPhys.87.1} {\bibfield
			{journal} {\bibinfo  {journal} {Rev. Mod. Phys.}\ }\textbf {\bibinfo {volume}
				{87}},\ \bibinfo {pages} {1} (\bibinfo {year} {2015})}\BibitemShut {NoStop}%
		\bibitem [{\citenamefont {Altland}\ and\ \citenamefont
			{Zirnbauer}(1997)}]{Altland1997}%
		\BibitemOpen
		\bibfield  {author} {\bibinfo {author} {\bibfnamefont {A.}~\bibnamefont
				{Altland}}\ and\ \bibinfo {author} {\bibfnamefont {M.~R.}\ \bibnamefont
				{Zirnbauer}},\ }\href {https://doi.org/10.1103/PhysRevB.55.1142} {\bibfield
			{journal} {\bibinfo  {journal} {Phys. Rev. B}\ }\textbf {\bibinfo {volume}
				{55}},\ \bibinfo {pages} {1142} (\bibinfo {year} {1997})}\BibitemShut
		{NoStop}%
		\bibitem [{\citenamefont {Ryu}\ \emph {et~al.}(2010)\citenamefont {Ryu},
			\citenamefont {Schnyder}, \citenamefont {Furusaki},\ and\ \citenamefont
			{Ludwig}}]{Ryu2010}%
		\BibitemOpen
		\bibfield  {author} {\bibinfo {author} {\bibfnamefont {S.}~\bibnamefont
				{Ryu}}, \bibinfo {author} {\bibfnamefont {A.~P.}\ \bibnamefont {Schnyder}},
			\bibinfo {author} {\bibfnamefont {A.}~\bibnamefont {Furusaki}},\ and\
			\bibinfo {author} {\bibfnamefont {A.~W.~W.}\ \bibnamefont {Ludwig}},\ }\href
		{https://doi.org/10.1088/1367-2630/12/6/065010} {\bibfield  {journal}
			{\bibinfo  {journal} {New Journal of Physics}\ }\textbf {\bibinfo {volume}
				{12}},\ \bibinfo {pages} {065010} (\bibinfo {year} {2010})}\BibitemShut
		{NoStop}%
		\bibitem [{\citenamefont {Wu}\ \emph {et~al.}(2007)\citenamefont {Wu},
			\citenamefont {Bergman}, \citenamefont {Balents},\ and\ \citenamefont
			{Das~Sarma}}]{Wu2007}%
		\BibitemOpen
		\bibfield  {author} {\bibinfo {author} {\bibfnamefont {C.}~\bibnamefont
				{Wu}}, \bibinfo {author} {\bibfnamefont {D.}~\bibnamefont {Bergman}},
			\bibinfo {author} {\bibfnamefont {L.}~\bibnamefont {Balents}},\ and\ \bibinfo
			{author} {\bibfnamefont {S.}~\bibnamefont {Das~Sarma}},\ }\href
		{https://doi.org/10.1103/PhysRevLett.99.070401} {\bibfield  {journal}
			{\bibinfo  {journal} {Phys. Rev. Lett.}\ }\textbf {\bibinfo {volume} {99}},\
			\bibinfo {pages} {070401} (\bibinfo {year} {2007})}\BibitemShut {NoStop}%
		\bibitem [{\citenamefont {Chen}\ and\ \citenamefont {Xie}(2019)}]{Chen2019}%
		\BibitemOpen
		\bibfield  {author} {\bibinfo {author} {\bibfnamefont {H.}~\bibnamefont
				{Chen}}\ and\ \bibinfo {author} {\bibfnamefont {X.~C.}\ \bibnamefont {Xie}},\
		}\href {https://doi.org/10.1103/PhysRevA.100.013601} {\bibfield  {journal}
			{\bibinfo  {journal} {Phys. Rev. A}\ }\textbf {\bibinfo {volume} {100}},\
			\bibinfo {pages} {013601} (\bibinfo {year} {2019})}\BibitemShut {NoStop}%
		\bibitem [{\citenamefont {Wu}\ and\ \citenamefont {Das~Sarma}(2008)}]{Wu2008}%
		\BibitemOpen
		\bibfield  {author} {\bibinfo {author} {\bibfnamefont {C.}~\bibnamefont
				{Wu}}\ and\ \bibinfo {author} {\bibfnamefont {S.}~\bibnamefont {Das~Sarma}},\
		}\href {https://doi.org/10.1103/PhysRevB.77.235107} {\bibfield  {journal}
			{\bibinfo  {journal} {Phys. Rev. B}\ }\textbf {\bibinfo {volume} {77}},\
			\bibinfo {pages} {235107} (\bibinfo {year} {2008})}\BibitemShut {NoStop}%
		\bibitem [{\citenamefont {Volovik}(2009)}]{Volovik2009}%
		\BibitemOpen
		\bibfield  {author} {\bibinfo {author} {\bibfnamefont {G.~E.}\ \bibnamefont
				{Volovik}},\ }\href@noop {} {\emph {\bibinfo {title} {The Universe in a
					Helium Droplet}}}\ (\bibinfo  {publisher} {OUP Oxford},\ \bibinfo {address}
		{Oxford},\ \bibinfo {year} {2009})\BibitemShut {NoStop}%
		\bibitem [{\citenamefont {Pines}\ and\ \citenamefont
			{Nozières}(2018)}]{Pines2018}%
		\BibitemOpen
		\bibfield  {author} {\bibinfo {author} {\bibfnamefont {D.}~\bibnamefont
				{Pines}}\ and\ \bibinfo {author} {\bibfnamefont {P.}~\bibnamefont
				{Nozières}},\ }\href {https://doi.org/10.4324/9780429492662} {\emph
			{\bibinfo {title} {The Theory of Quantum Liquids: Normal Fermi Liquids}}}\
		(\bibinfo  {publisher} {CRC Press},\ \bibinfo {year} {2018})\BibitemShut
		{NoStop}%
		\bibitem [{\citenamefont {Klitzing}\ \emph {et~al.}(1980)\citenamefont
			{Klitzing}, \citenamefont {Dorda},\ and\ \citenamefont
			{Pepper}}]{Klitzing1980}%
		\BibitemOpen
		\bibfield  {author} {\bibinfo {author} {\bibfnamefont {K.~v.}\ \bibnamefont
				{Klitzing}}, \bibinfo {author} {\bibfnamefont {G.}~\bibnamefont {Dorda}},\
			and\ \bibinfo {author} {\bibfnamefont {M.}~\bibnamefont {Pepper}},\ }\href
		{https://doi.org/10.1103/PhysRevLett.45.494} {\bibfield  {journal} {\bibinfo
				{journal} {Phys. Rev. Lett.}\ }\textbf {\bibinfo {volume} {45}},\ \bibinfo
			{pages} {494} (\bibinfo {year} {1980})}\BibitemShut {NoStop}%
		\bibitem [{\citenamefont {Thouless}\ \emph {et~al.}(1982)\citenamefont
			{Thouless}, \citenamefont {Kohmoto}, \citenamefont {Nightingale},\ and\
			\citenamefont {den Nijs}}]{Thouless1982}%
		\BibitemOpen
		\bibfield  {author} {\bibinfo {author} {\bibfnamefont {D.~J.}\ \bibnamefont
				{Thouless}}, \bibinfo {author} {\bibfnamefont {M.}~\bibnamefont {Kohmoto}},
			\bibinfo {author} {\bibfnamefont {M.~P.}\ \bibnamefont {Nightingale}},\ and\
			\bibinfo {author} {\bibfnamefont {M.}~\bibnamefont {den Nijs}},\ }\href
		{https://doi.org/10.1103/PhysRevLett.49.405} {\bibfield  {journal} {\bibinfo
				{journal} {Phys. Rev. Lett.}\ }\textbf {\bibinfo {volume} {49}},\ \bibinfo
			{pages} {405} (\bibinfo {year} {1982})}\BibitemShut {NoStop}%
		\bibitem [{\citenamefont {Kohmoto}(1985)}]{Kohmoto1985}%
		\BibitemOpen
		\bibfield  {author} {\bibinfo {author} {\bibfnamefont {M.}~\bibnamefont
				{Kohmoto}},\ }\href {https://doi.org/10.1016/0003-4916(85)90148-4} {\bibfield
			{journal} {\bibinfo  {journal} {Annals of Physics}\ }\textbf {\bibinfo
				{volume} {160}},\ \bibinfo {pages} {343} (\bibinfo {year}
			{1985})}\BibitemShut {NoStop}%
		\bibitem [{\citenamefont {MacDonald}\ \emph {et~al.}(1988)\citenamefont
			{MacDonald}, \citenamefont {Girvin},\ and\ \citenamefont
			{Yoshioka}}]{MacDonald1988}%
		\BibitemOpen
		\bibfield  {author} {\bibinfo {author} {\bibfnamefont {A.~H.}\ \bibnamefont
				{MacDonald}}, \bibinfo {author} {\bibfnamefont {S.~M.}\ \bibnamefont
				{Girvin}},\ and\ \bibinfo {author} {\bibfnamefont {D.}~\bibnamefont
				{Yoshioka}},\ }\href {https://doi.org/10.1103/PhysRevB.37.9753} {\bibfield
			{journal} {\bibinfo  {journal} {Phys. Rev. B}\ }\textbf {\bibinfo {volume}
				{37}},\ \bibinfo {pages} {9753} (\bibinfo {year} {1988})}\BibitemShut
		{NoStop}%
		\bibitem [{\citenamefont {Fazekas}(1999)}]{Fazekas1999}%
		\BibitemOpen
		\bibfield  {author} {\bibinfo {author} {\bibfnamefont {P.}~\bibnamefont
				{Fazekas}},\ }\href {https://doi.org/10.1142/2945} {\emph {\bibinfo {title}
				{Lecture Notes on Electron Correlation and Magnetism}}}\ (\bibinfo
		{publisher} {WORLD SCIENTIFIC},\ \bibinfo {year} {1999})\BibitemShut
		{NoStop}%
		\bibitem [{\citenamefont {Kuklov}\ and\ \citenamefont
			{Svistunov}(2003)}]{Kuklov2003}%
		\BibitemOpen
		\bibfield  {author} {\bibinfo {author} {\bibfnamefont {A.~B.}\ \bibnamefont
				{Kuklov}}\ and\ \bibinfo {author} {\bibfnamefont {B.~V.}\ \bibnamefont
				{Svistunov}},\ }\href {https://doi.org/10.1103/PhysRevLett.90.100401}
		{\bibfield  {journal} {\bibinfo  {journal} {Phys. Rev. Lett.}\ }\textbf
			{\bibinfo {volume} {90}},\ \bibinfo {pages} {100401} (\bibinfo {year}
			{2003})}\BibitemShut {NoStop}%
		\bibitem [{\citenamefont {Luttinger}\ and\ \citenamefont
			{Tisza}(1946)}]{Luttinger1946}%
		\BibitemOpen
		\bibfield  {author} {\bibinfo {author} {\bibfnamefont {J.~M.}\ \bibnamefont
				{Luttinger}}\ and\ \bibinfo {author} {\bibfnamefont {L.}~\bibnamefont
				{Tisza}},\ }\href {https://doi.org/10.1103/PhysRev.70.954} {\bibfield
			{journal} {\bibinfo  {journal} {Phys. Rev.}\ }\textbf {\bibinfo {volume}
				{70}},\ \bibinfo {pages} {954} (\bibinfo {year} {1946})}\BibitemShut
		{NoStop}%
		\bibitem [{\citenamefont {Holstein}\ and\ \citenamefont
			{Primakoff}(1940)}]{Holstein1940}%
		\BibitemOpen
		\bibfield  {author} {\bibinfo {author} {\bibfnamefont {T.}~\bibnamefont
				{Holstein}}\ and\ \bibinfo {author} {\bibfnamefont {H.}~\bibnamefont
				{Primakoff}},\ }\href {https://doi.org/10.1103/PhysRev.58.1098} {\bibfield
			{journal} {\bibinfo  {journal} {Phys. Rev.}\ }\textbf {\bibinfo {volume}
				{58}},\ \bibinfo {pages} {1098} (\bibinfo {year} {1940})}\BibitemShut
		{NoStop}%
		\bibitem [{\citenamefont {Villain}(1977)}]{Villain1977}%
		\BibitemOpen
		\bibfield  {author} {\bibinfo {author} {\bibfnamefont {J.}~\bibnamefont
				{Villain}},\ }\href {https://doi.org/10.1051/jphys:01977003804038500}
		{\bibfield  {journal} {\bibinfo  {journal} {Journal de Physique}\ }\textbf
			{\bibinfo {volume} {38}},\ \bibinfo {pages} {385} (\bibinfo {year}
			{1977})}\BibitemShut {NoStop}%
		\bibitem [{\citenamefont {Villain}\ \emph {et~al.}(1980)\citenamefont
			{Villain}, \citenamefont {Bidaux}, \citenamefont {Carton},\ and\
			\citenamefont {Conte}}]{Villain1980}%
		\BibitemOpen
		\bibfield  {author} {\bibinfo {author} {\bibfnamefont {J.}~\bibnamefont
				{Villain}}, \bibinfo {author} {\bibfnamefont {R.}~\bibnamefont {Bidaux}},
			\bibinfo {author} {\bibfnamefont {J.-P.}\ \bibnamefont {Carton}},\ and\
			\bibinfo {author} {\bibfnamefont {R.}~\bibnamefont {Conte}},\ }\href
		{https://doi.org/10.1051/jphys:0198000410110126300} {\bibfield  {journal}
			{\bibinfo  {journal} {Journal de Physique}\ }\textbf {\bibinfo {volume}
				{41}},\ \bibinfo {pages} {1263} (\bibinfo {year} {1980})}\BibitemShut
		{NoStop}%
		\bibitem [{\citenamefont {{Shender}}(1982)}]{Shender1982}%
		\BibitemOpen
		\bibfield  {author} {\bibinfo {author} {\bibfnamefont {E.~F.}\ \bibnamefont
				{{Shender}}},\ }\href@noop {} {\bibfield  {journal} {\bibinfo  {journal}
				{Soviet Journal of Experimental and Theoretical Physics}\ }\textbf {\bibinfo
				{volume} {56}},\ \bibinfo {pages} {178} (\bibinfo {year} {1982})}\BibitemShut
		{NoStop}%
		\bibitem [{\citenamefont {Henley}(1989)}]{Henley1989}%
		\BibitemOpen
		\bibfield  {author} {\bibinfo {author} {\bibfnamefont {C.~L.}\ \bibnamefont
				{Henley}},\ }\href {https://doi.org/10.1103/PhysRevLett.62.2056} {\bibfield
			{journal} {\bibinfo  {journal} {Phys. Rev. Lett.}\ }\textbf {\bibinfo
				{volume} {62}},\ \bibinfo {pages} {2056} (\bibinfo {year}
			{1989})}\BibitemShut {NoStop}%
		\bibitem [{\citenamefont {Claudine~Lacroix}(2011)}]{Lacroix2011}%
		\BibitemOpen
		\bibinfo {editor} {\bibfnamefont {F.~M.}\ \bibnamefont {Claudine~Lacroix},
			\bibfnamefont {Philippe~Mendels}},\ ed.,\ \href
		{https://doi.org/10.1007/978-3-642-10589-0} {\emph {\bibinfo {title}
				{Introduction to Frustrated Magnetism}}}\ (\bibinfo  {publisher} {Springer
			Berlin Heidelberg},\ \bibinfo {year} {2011})\BibitemShut {NoStop}%
		\bibitem [{\citenamefont {Diep}(2012)}]{Diep2012}%
		\BibitemOpen
		\bibfield  {author} {\bibinfo {author} {\bibfnamefont {H.~T.}\ \bibnamefont
				{Diep}},\ }\href {https://doi.org/10.1142/8676} {\emph {\bibinfo {title}
				{Frustrated Spin Systems}}}\ (\bibinfo  {publisher} {{WORLD} {SCIENTIFIC}},\
		\bibinfo {year} {2012})\BibitemShut {NoStop}%
		\bibitem [{\citenamefont {Green}\ \emph {et~al.}(2018)\citenamefont {Green},
			\citenamefont {Conduit},\ and\ \citenamefont {Krüger}}]{Green2018}%
		\BibitemOpen
		\bibfield  {author} {\bibinfo {author} {\bibfnamefont {A.~G.}\ \bibnamefont
				{Green}}, \bibinfo {author} {\bibfnamefont {G.}~\bibnamefont {Conduit}},\
			and\ \bibinfo {author} {\bibfnamefont {F.}~\bibnamefont {Krüger}},\ }\href
		{https://doi.org/10.1146/annurev-conmatphys-033117-053925} {\bibfield
			{journal} {\bibinfo  {journal} {Annual Review of Condensed Matter Physics}\
			}\textbf {\bibinfo {volume} {9}},\ \bibinfo {pages} {59} (\bibinfo {year}
			{2018})}\BibitemShut {NoStop}%
		\bibitem [{\citenamefont {M\"uller}\ \emph {et~al.}(2007)\citenamefont
			{M\"uller}, \citenamefont {F\"olling}, \citenamefont {Widera},\ and\
			\citenamefont {Bloch}}]{Mueller2007}%
		\BibitemOpen
		\bibfield  {author} {\bibinfo {author} {\bibfnamefont {T.}~\bibnamefont
				{M\"uller}}, \bibinfo {author} {\bibfnamefont {S.}~\bibnamefont {F\"olling}},
			\bibinfo {author} {\bibfnamefont {A.}~\bibnamefont {Widera}},\ and\ \bibinfo
			{author} {\bibfnamefont {I.}~\bibnamefont {Bloch}},\ }\href
		{https://doi.org/10.1103/PhysRevLett.99.200405} {\bibfield  {journal}
			{\bibinfo  {journal} {Phys. Rev. Lett.}\ }\textbf {\bibinfo {volume} {99}},\
			\bibinfo {pages} {200405} (\bibinfo {year} {2007})}\BibitemShut {NoStop}%
		\bibitem [{\citenamefont {Wirth}\ \emph {et~al.}(2010)\citenamefont {Wirth},
			\citenamefont {Ölschläger},\ and\ \citenamefont {Hemmerich}}]{Wirth2010}%
		\BibitemOpen
		\bibfield  {author} {\bibinfo {author} {\bibfnamefont {G.}~\bibnamefont
				{Wirth}}, \bibinfo {author} {\bibfnamefont {M.}~\bibnamefont
				{Ölschläger}},\ and\ \bibinfo {author} {\bibfnamefont {A.}~\bibnamefont
				{Hemmerich}},\ }\href {https://doi.org/10.1038/nphys1857} {\bibfield
			{journal} {\bibinfo  {journal} {Nature Physics}\ }\textbf {\bibinfo {volume}
				{7}},\ \bibinfo {pages} {147} (\bibinfo {year} {2010})}\BibitemShut {NoStop}%
		\bibitem [{\citenamefont {Soltan-Panahi}\ \emph {et~al.}(2011)\citenamefont
			{Soltan-Panahi}, \citenamefont {Lühmann}, \citenamefont {Struck},
			\citenamefont {Windpassinger},\ and\ \citenamefont
			{Sengstock}}]{SoltanPanahi2011}%
		\BibitemOpen
		\bibfield  {author} {\bibinfo {author} {\bibfnamefont {P.}~\bibnamefont
				{Soltan-Panahi}}, \bibinfo {author} {\bibfnamefont {D.-S.}\ \bibnamefont
				{Lühmann}}, \bibinfo {author} {\bibfnamefont {J.}~\bibnamefont {Struck}},
			\bibinfo {author} {\bibfnamefont {P.}~\bibnamefont {Windpassinger}},\ and\
			\bibinfo {author} {\bibfnamefont {K.}~\bibnamefont {Sengstock}},\ }\href
		{https://doi.org/10.1038/nphys2128} {\bibfield  {journal} {\bibinfo
				{journal} {Nature Physics}\ }\textbf {\bibinfo {volume} {8}},\ \bibinfo
			{pages} {71} (\bibinfo {year} {2011})}\BibitemShut {NoStop}%
		\bibitem [{\citenamefont {Ölschläger}\ \emph {et~al.}(2013)\citenamefont
			{Ölschläger}, \citenamefont {Kock}, \citenamefont {Wirth}, \citenamefont
			{Ewerbeck}, \citenamefont {Morais~Smith},\ and\ \citenamefont
			{Hemmerich}}]{Oelschlaeger2013}%
		\BibitemOpen
		\bibfield  {author} {\bibinfo {author} {\bibfnamefont {M.}~\bibnamefont
				{Ölschläger}}, \bibinfo {author} {\bibfnamefont {T.}~\bibnamefont {Kock}},
			\bibinfo {author} {\bibfnamefont {G.}~\bibnamefont {Wirth}}, \bibinfo
			{author} {\bibfnamefont {A.}~\bibnamefont {Ewerbeck}}, \bibinfo {author}
			{\bibfnamefont {C.}~\bibnamefont {Morais~Smith}},\ and\ \bibinfo {author}
			{\bibfnamefont {A.}~\bibnamefont {Hemmerich}},\ }\href
		{https://doi.org/10.1088/1367-2630/15/8/083041} {\bibfield  {journal}
			{\bibinfo  {journal} {New Journal of Physics}\ }\textbf {\bibinfo {volume}
				{15}},\ \bibinfo {pages} {083041} (\bibinfo {year} {2013})}\BibitemShut
		{NoStop}%
		\bibitem [{\citenamefont {Kock}\ \emph {et~al.}(2015)\citenamefont {Kock},
			\citenamefont {\"Olschl\"ager}, \citenamefont {Ewerbeck}, \citenamefont
			{Huang}, \citenamefont {Mathey},\ and\ \citenamefont {Hemmerich}}]{Kock2015}%
		\BibitemOpen
		\bibfield  {author} {\bibinfo {author} {\bibfnamefont {T.}~\bibnamefont
				{Kock}}, \bibinfo {author} {\bibfnamefont {M.}~\bibnamefont
				{\"Olschl\"ager}}, \bibinfo {author} {\bibfnamefont {A.}~\bibnamefont
				{Ewerbeck}}, \bibinfo {author} {\bibfnamefont {W.-M.}\ \bibnamefont {Huang}},
			\bibinfo {author} {\bibfnamefont {L.}~\bibnamefont {Mathey}},\ and\ \bibinfo
			{author} {\bibfnamefont {A.}~\bibnamefont {Hemmerich}},\ }\href
		{https://doi.org/10.1103/PhysRevLett.114.115301} {\bibfield  {journal}
			{\bibinfo  {journal} {Phys. Rev. Lett.}\ }\textbf {\bibinfo {volume} {114}},\
			\bibinfo {pages} {115301} (\bibinfo {year} {2015})}\BibitemShut {NoStop}%
		\bibitem [{\citenamefont {Jin}\ \emph {et~al.}(2021)\citenamefont {Jin},
			\citenamefont {Zhang}, \citenamefont {Guo}, \citenamefont {Chen},
			\citenamefont {Zhou},\ and\ \citenamefont {Li}}]{Jin2021}%
		\BibitemOpen
		\bibfield  {author} {\bibinfo {author} {\bibfnamefont {S.}~\bibnamefont
				{Jin}}, \bibinfo {author} {\bibfnamefont {W.}~\bibnamefont {Zhang}}, \bibinfo
			{author} {\bibfnamefont {X.}~\bibnamefont {Guo}}, \bibinfo {author}
			{\bibfnamefont {X.}~\bibnamefont {Chen}}, \bibinfo {author} {\bibfnamefont
				{X.}~\bibnamefont {Zhou}},\ and\ \bibinfo {author} {\bibfnamefont
				{X.}~\bibnamefont {Li}},\ }\href
		{https://doi.org/10.1103/PhysRevLett.126.035301} {\bibfield  {journal}
			{\bibinfo  {journal} {Phys. Rev. Lett.}\ }\textbf {\bibinfo {volume} {126}},\
			\bibinfo {pages} {035301} (\bibinfo {year} {2021})}\BibitemShut {NoStop}%
		\bibitem [{\citenamefont {Vargas}\ \emph {et~al.}(2021)\citenamefont {Vargas},
			\citenamefont {Nuske}, \citenamefont {Eichberger}, \citenamefont {Hippler},
			\citenamefont {Mathey},\ and\ \citenamefont {Hemmerich}}]{Vargas2021}%
		\BibitemOpen
		\bibfield  {author} {\bibinfo {author} {\bibfnamefont {J.}~\bibnamefont
				{Vargas}}, \bibinfo {author} {\bibfnamefont {M.}~\bibnamefont {Nuske}},
			\bibinfo {author} {\bibfnamefont {R.}~\bibnamefont {Eichberger}}, \bibinfo
			{author} {\bibfnamefont {C.}~\bibnamefont {Hippler}}, \bibinfo {author}
			{\bibfnamefont {L.}~\bibnamefont {Mathey}},\ and\ \bibinfo {author}
			{\bibfnamefont {A.}~\bibnamefont {Hemmerich}},\ }\href
		{https://doi.org/10.1103/PhysRevLett.126.200402} {\bibfield  {journal}
			{\bibinfo  {journal} {Phys. Rev. Lett.}\ }\textbf {\bibinfo {volume} {126}},\
			\bibinfo {pages} {200402} (\bibinfo {year} {2021})}\BibitemShut {NoStop}%
		\bibitem [{\citenamefont {Wang}\ \emph {et~al.}(2021)\citenamefont {Wang},
			\citenamefont {Luo}, \citenamefont {Liu}, \citenamefont {Liu}, \citenamefont
			{Hemmerich},\ and\ \citenamefont {Xu}}]{Wang2021}%
		\BibitemOpen
		\bibfield  {author} {\bibinfo {author} {\bibfnamefont {X.-Q.}\ \bibnamefont
				{Wang}}, \bibinfo {author} {\bibfnamefont {G.-Q.}\ \bibnamefont {Luo}},
			\bibinfo {author} {\bibfnamefont {J.-Y.}\ \bibnamefont {Liu}}, \bibinfo
			{author} {\bibfnamefont {W.~V.}\ \bibnamefont {Liu}}, \bibinfo {author}
			{\bibfnamefont {A.}~\bibnamefont {Hemmerich}},\ and\ \bibinfo {author}
			{\bibfnamefont {Z.-F.}\ \bibnamefont {Xu}},\ }\href
		{https://doi.org/10.1038/s41586-021-03702-0} {\bibfield  {journal} {\bibinfo
				{journal} {Nature}\ }\textbf {\bibinfo {volume} {596}},\ \bibinfo {pages}
			{227} (\bibinfo {year} {2021})}\BibitemShut {NoStop}%
		\bibitem [{\citenamefont {Wang}\ \emph {et~al.}(2023)\citenamefont {Wang},
			\citenamefont {Luo}, \citenamefont {Liu}, \citenamefont {Huang},
			\citenamefont {Li}, \citenamefont {Wu}, \citenamefont {Hemmerich},\ and\
			\citenamefont {Xu}}]{Wang2023}%
		\BibitemOpen
		\bibfield  {author} {\bibinfo {author} {\bibfnamefont {X.-Q.}\ \bibnamefont
				{Wang}}, \bibinfo {author} {\bibfnamefont {G.-Q.}\ \bibnamefont {Luo}},
			\bibinfo {author} {\bibfnamefont {J.-Y.}\ \bibnamefont {Liu}}, \bibinfo
			{author} {\bibfnamefont {G.-H.}\ \bibnamefont {Huang}}, \bibinfo {author}
			{\bibfnamefont {Z.-X.}\ \bibnamefont {Li}}, \bibinfo {author} {\bibfnamefont
				{C.}~\bibnamefont {Wu}}, \bibinfo {author} {\bibfnamefont {A.}~\bibnamefont
				{Hemmerich}},\ and\ \bibinfo {author} {\bibfnamefont {Z.-F.}\ \bibnamefont
				{Xu}},\ }\href {https://doi.org/10.1103/PhysRevLett.131.226001} {\bibfield
			{journal} {\bibinfo  {journal} {Phys. Rev. Lett.}\ }\textbf {\bibinfo
				{volume} {131}},\ \bibinfo {pages} {226001} (\bibinfo {year}
			{2023})}\BibitemShut {NoStop}%
		\bibitem [{\citenamefont {K\"ohl}\ \emph {et~al.}(2005)\citenamefont {K\"ohl},
			\citenamefont {Moritz}, \citenamefont {St\"oferle}, \citenamefont
			{G\"unter},\ and\ \citenamefont {Esslinger}}]{Koehl2005}%
		\BibitemOpen
		\bibfield  {author} {\bibinfo {author} {\bibfnamefont {M.}~\bibnamefont
				{K\"ohl}}, \bibinfo {author} {\bibfnamefont {H.}~\bibnamefont {Moritz}},
			\bibinfo {author} {\bibfnamefont {T.}~\bibnamefont {St\"oferle}}, \bibinfo
			{author} {\bibfnamefont {K.}~\bibnamefont {G\"unter}},\ and\ \bibinfo
			{author} {\bibfnamefont {T.}~\bibnamefont {Esslinger}},\ }\href
		{https://doi.org/10.1103/PhysRevLett.94.080403} {\bibfield  {journal}
			{\bibinfo  {journal} {Phys. Rev. Lett.}\ }\textbf {\bibinfo {volume} {94}},\
			\bibinfo {pages} {080403} (\bibinfo {year} {2005})}\BibitemShut {NoStop}%
		\bibitem [{\citenamefont {Mamaev}\ \emph {et~al.}(2021)\citenamefont {Mamaev},
			\citenamefont {He}, \citenamefont {Bilitewski}, \citenamefont {Venu},
			\citenamefont {Thywissen},\ and\ \citenamefont {Rey}}]{Mamaev2021}%
		\BibitemOpen
		\bibfield  {author} {\bibinfo {author} {\bibfnamefont {M.}~\bibnamefont
				{Mamaev}}, \bibinfo {author} {\bibfnamefont {P.}~\bibnamefont {He}}, \bibinfo
			{author} {\bibfnamefont {T.}~\bibnamefont {Bilitewski}}, \bibinfo {author}
			{\bibfnamefont {V.}~\bibnamefont {Venu}}, \bibinfo {author} {\bibfnamefont
				{J.~H.}\ \bibnamefont {Thywissen}},\ and\ \bibinfo {author} {\bibfnamefont
				{A.~M.}\ \bibnamefont {Rey}},\ }\href
		{https://doi.org/10.1103/PhysRevLett.127.143401} {\bibfield  {journal}
			{\bibinfo  {journal} {Phys. Rev. Lett.}\ }\textbf {\bibinfo {volume} {127}},\
			\bibinfo {pages} {143401} (\bibinfo {year} {2021})}\BibitemShut {NoStop}%
		\bibitem [{\citenamefont {Venu}\ \emph {et~al.}(2023)\citenamefont {Venu},
			\citenamefont {Xu}, \citenamefont {Mamaev}, \citenamefont {Corapi},
			\citenamefont {Bilitewski}, \citenamefont {D’Incao}, \citenamefont
			{Fujiwara}, \citenamefont {Rey},\ and\ \citenamefont {Thywissen}}]{Venu2023}%
		\BibitemOpen
		\bibfield  {author} {\bibinfo {author} {\bibfnamefont {V.}~\bibnamefont
				{Venu}}, \bibinfo {author} {\bibfnamefont {P.}~\bibnamefont {Xu}}, \bibinfo
			{author} {\bibfnamefont {M.}~\bibnamefont {Mamaev}}, \bibinfo {author}
			{\bibfnamefont {F.}~\bibnamefont {Corapi}}, \bibinfo {author} {\bibfnamefont
				{T.}~\bibnamefont {Bilitewski}}, \bibinfo {author} {\bibfnamefont {J.~P.}\
				\bibnamefont {D’Incao}}, \bibinfo {author} {\bibfnamefont {C.~J.}\
				\bibnamefont {Fujiwara}}, \bibinfo {author} {\bibfnamefont {A.~M.}\
				\bibnamefont {Rey}},\ and\ \bibinfo {author} {\bibfnamefont {J.~H.}\
				\bibnamefont {Thywissen}},\ }\href
		{https://doi.org/10.1038/s41586-022-05405-6} {\bibfield  {journal} {\bibinfo
				{journal} {Nature}\ }\textbf {\bibinfo {volume} {613}},\ \bibinfo {pages}
			{262} (\bibinfo {year} {2023})}\BibitemShut {NoStop}%
		\bibitem [{\citenamefont {Nagano}\ \emph {et~al.}(2007)\citenamefont {Nagano},
			\citenamefont {Naka}, \citenamefont {Nasu},\ and\ \citenamefont
			{Ishihara}}]{Nagano2007}%
		\BibitemOpen
		\bibfield  {author} {\bibinfo {author} {\bibfnamefont {A.}~\bibnamefont
				{Nagano}}, \bibinfo {author} {\bibfnamefont {M.}~\bibnamefont {Naka}},
			\bibinfo {author} {\bibfnamefont {J.}~\bibnamefont {Nasu}},\ and\ \bibinfo
			{author} {\bibfnamefont {S.}~\bibnamefont {Ishihara}},\ }\href
		{https://doi.org/10.1103/PhysRevLett.99.217202} {\bibfield  {journal}
			{\bibinfo  {journal} {Phys. Rev. Lett.}\ }\textbf {\bibinfo {volume} {99}},\
			\bibinfo {pages} {217202} (\bibinfo {year} {2007})}\BibitemShut {NoStop}%
		\bibitem [{\citenamefont {Nasu}\ \emph {et~al.}(2008)\citenamefont {Nasu},
			\citenamefont {Nagano}, \citenamefont {Naka},\ and\ \citenamefont
			{Ishihara}}]{Nasu2008}%
		\BibitemOpen
		\bibfield  {author} {\bibinfo {author} {\bibfnamefont {J.}~\bibnamefont
				{Nasu}}, \bibinfo {author} {\bibfnamefont {A.}~\bibnamefont {Nagano}},
			\bibinfo {author} {\bibfnamefont {M.}~\bibnamefont {Naka}},\ and\ \bibinfo
			{author} {\bibfnamefont {S.}~\bibnamefont {Ishihara}},\ }\href
		{https://doi.org/10.1103/PhysRevB.78.024416} {\bibfield  {journal} {\bibinfo
				{journal} {Phys. Rev. B}\ }\textbf {\bibinfo {volume} {78}},\ \bibinfo
			{pages} {024416} (\bibinfo {year} {2008})}\BibitemShut {NoStop}%
		\bibitem [{\citenamefont {Zhao}\ and\ \citenamefont {Liu}(2008)}]{Zhao2008}%
		\BibitemOpen
		\bibfield  {author} {\bibinfo {author} {\bibfnamefont {E.}~\bibnamefont
				{Zhao}}\ and\ \bibinfo {author} {\bibfnamefont {W.~V.}\ \bibnamefont {Liu}},\
		}\href {https://doi.org/10.1103/PhysRevLett.100.160403} {\bibfield  {journal}
			{\bibinfo  {journal} {Phys. Rev. Lett.}\ }\textbf {\bibinfo {volume} {100}},\
			\bibinfo {pages} {160403} (\bibinfo {year} {2008})}\BibitemShut {NoStop}%
		\bibitem [{\citenamefont {Wu}(2008)}]{Wu2008b}%
		\BibitemOpen
		\bibfield  {author} {\bibinfo {author} {\bibfnamefont {C.}~\bibnamefont
				{Wu}},\ }\href {https://doi.org/10.1103/PhysRevLett.100.200406} {\bibfield
			{journal} {\bibinfo  {journal} {Phys. Rev. Lett.}\ }\textbf {\bibinfo
				{volume} {100}},\ \bibinfo {pages} {200406} (\bibinfo {year}
			{2008})}\BibitemShut {NoStop}%
		\bibitem [{\citenamefont {Blaizot}\ and\ \citenamefont
			{Ripka}(1986)}]{Blaizot1986}%
		\BibitemOpen
		\bibfield  {author} {\bibinfo {author} {\bibfnamefont {J.-P.}\ \bibnamefont
				{Blaizot}}\ and\ \bibinfo {author} {\bibfnamefont {G.}~\bibnamefont
				{Ripka}},\ }\href@noop {} {\emph {\bibinfo {title} {Quantum theory of finite
					systems}}}\ (\bibinfo  {publisher} {MIT Press},\ \bibinfo {address}
		{Cambridge, Mass.},\ \bibinfo {year} {1986})\BibitemShut {NoStop}%
	\end{thebibliography}
	
	%

	\clearpage

	\widetext
	
	\begin{center}
		\textbf{\large Supplemental Material for "Interaction-Driven Intervalley Coherence with Emergent Kekul\'e Orbitons"}
	\end{center}
	
		
	\section{Orbital exchange Hamiltonian}
	\label{app:OEX}
	
	\begin{table*}[b]
		\caption{\label{tab:gamma-s} 
			The $i$-th eigenstate $\Gamma_n^i$ of the Hamiltonian $H_\text{I}$ in Eq.~(\ref{eq:HIS}) with the corresponding eigenenergy $E_{\Gamma_n^i}$ for $p^{n=0,1,2}$ configurations. Here $\left|\text{vac}\right\rangle$ is the vacuum state.}
		\begin{ruledtabular}
			\begin{tabular}{cccccccc}
				&\multicolumn{1}{c}{$p^0$ configuration}&\multicolumn{2}{c}{$p^1$ configuration}&\multicolumn{1}{c}{$p^2$ configuration}\\ \hline
				$i$	
				&	$1$	
				&	$1$	&	$2$
				&	$1$		\\ 
				$E_{\Gamma_n^i}$	
				&	$0$	
				&	$0$	&	$0$
				&	$U$		\\ 	
				$\left|\Gamma_n^i\right\rangle$	
				&	$\left|\text{vac}\right\rangle$	
				&	$p_{x}^\dagger\left|\text{vac}\right\rangle$	
				&	$p_{y}^\dagger\left|\text{vac}\right\rangle$	
				&	$p_{x}^\dagger p_{y}^\dagger\left|\text{vac}\right\rangle$	\\					
			\end{tabular}
		\end{ruledtabular}
	\end{table*}
	
	To derive the effective low-energy Hamiltonian, we shall first diagonalize the on-site Hubbard interaction 
	\begin{equation}
		H_\text{I}=U\hat{n}_x\hat{n}_y 
		=\sum_{\Gamma_n}E_{\Gamma_n^i}\left|\Gamma_n^i\right\rangle\left\langle\Gamma_n^i\right|
		\label{eq:HIS}
	\end{equation}
	where $\hat{n}_\mu=p^\dagger_\mu p_\mu$ is the density operator and the explicit eigenstates $\Gamma_n^i$ with the corresponding eigenenergies $E_{\Gamma_n^i}$ for $p^{n=0,1,2}$ configurations are listed in Table.~\ref{tab:gamma-s}.
		
	Next, the orbital exchange model in the Mott insulating phase $n=1$ is constructed through the degenerate perturbation theory. The $p^1$ configuration with zero energy is an orbital doublet with one fermion occupying either $p_x$ or $p_y$ orbital. Note that the charge excitations $\left(p^1\right)_i\left(p^1\right)_j\rightleftharpoons\left(p^2\right)_i\left(p^0\right)_j$ and $\left(p^1\right)_i\left(p^1\right)_j\rightleftharpoons\left(p^0\right)_i\left(p^2\right)_j$ through the inter-site hopping processes between $i$-th and $j$-th sites have an energy gap that is proportional to the Hubbard interaction $U$. In the large-$U$ regime, the effective low-energy model is described by the second-order hopping process with both the initial and final states in $p^{1}$ configuration, which involves no charge gap. Let us derive the orbital exchange interaction along an isolated bond $\bm{e}$ connecting $i$-th and $j$-th sites. The inter-site hopping is described by
	\begin{equation}
		\hat{T}=
			\underset{\equiv T^+}{\underbrace{\sum_{\mu\nu} T_{\mu\nu}p^\dagger_{\mu i} p_{\nu j}}}		 
		+	\underset{\equiv T^-}{\underbrace{\sum_{\mu\nu} T_{\mu\nu}^* p^\dagger_{\nu j} p_{\mu i}}}, 
	\end{equation}
	with $T_{\mu\nu} = \left(t_\sigma-t_\pi\right)\hat{e}^\mu\hat{e}^\nu+\delta_{\mu,\nu}t_\pi$ being the bond-vector dependent hopping due to the anisotropy of $p$ orbitals. Employing the second-order perturbation theory~\cite{MacDonald1988,Fazekas1999,Kuklov2003}, the general matrix elements of orbital exchange interaction are given by
	\begin{equation}
		\mathbb{J}_{kl,k^\prime l^\prime}=\sum_{mn}
		\frac{
			\left\langle
			\underset{i\text{-th site}}{\Gamma_1^k}
			\underset{j\text{-th site}}{\Gamma_1^l}\right|
			T^+
			\left| \underset{i\text{-th site}}{\Gamma_2^m}
			\underset{j\text{-th site}}{\Gamma_0^n} 
			\right\rangle 
			\left\langle
			\underset{i\text{-th site}}{\Gamma_2^m}
			\underset{j\text{-th site}}{\Gamma_0^n}\right|
			T^-
			\left| \underset{i\text{-th site}}{\Gamma_1^{k^\prime}}
			\underset{j\text{-th site}}{\Gamma_1^{l^\prime}}
			\right\rangle
			}
			{-E_{\Gamma_{2}^m}-E_{\Gamma_0^n}} 
		+\left(i\leftrightarrow j\right). 
	\end{equation}
	To describe the orbital exchange Hamiltonian, we introduce the orbital pseudospin operators  
	$\{\tau_+,\tau_-\} \equiv \{p^\dagger_{x}p_{y},p^\dagger_{y}p_{x}\}$,
	which flip the states of the orbital doublet in $p^1$ configuration. The $z$ component of pseudospin $\bm{\tau}$ follows through the spin-$1/2$ angular momentum algebra $\tau_z=\left[\tau_+,\tau_-\right]$. 
	A lengthy but straightforward algebra leads to the orbital exchange Hamiltonian
	\begin{equation}
		H_\text{OE} =  
		J\left[\cos\left(2\theta\right)\tau_z^i+\sin\left(2\theta\right)\tau_x^i\right]
		\left[\cos\left(2\theta\right)\tau_z^j+\sin\left(2\theta\right)\tau_x^j\right]
		+ J^\prime\bm{\tau}^i\cdot\bm{\tau}^j
	\end{equation}
	with 
	\begin{eqnarray}
		\{J,J^\prime\}&=&\{\frac{\left(t_\sigma-t_\pi\right)^2}{2U},\frac{t_\sigma t_\pi}{U}\}. 
	\end{eqnarray}
	Here the bond vector is parameterized as $\hat{e}=\cos\theta\hat{x}+\sin\theta\hat{y}$.
	
	Finally, the total orbital exchange Hamiltonian by summing all bonds in the honeycomb lattice takes the form 
	\begin{eqnarray}
		H_\text{OE}=
		J\sum_{\langle ij \rangle\parallel\bm{e}_\gamma}\tau^i_\gamma\tau^{j}_\gamma
		+J^\prime \sum_{\langle ij \rangle}	\bm{\tau}^i\cdot\bm{\tau}^j
		\label{eq:OES}
	\end{eqnarray}
	with the bond dependent terms
	\begin{equation}
		\tau_\gamma=\tau_z\cos\left[2\theta_\gamma\right]+\tau_x\sin\left[2\theta_\gamma\right]
		\text{ and }
		\{\theta_1,\theta_2,\theta_3\}=\{\frac{1}{6}\pi,\frac{5}{6}\pi,\frac{3}{2}\pi\}.
	\end{equation}
	Unlike usual spin systems, the first term in the orbital exchange Eq.~(\ref{eq:OES}) is anisotropic. It roots in the fact that the hopping processes are bond dependent due to the spatial orientation of $p$ orbitals.  	
	
	\section{Classical ground-state construction}
	\label{app:CGS}
	
	The classical ground state is obtained by treating the orbital pseudospin $\bm{\tau}$ as a classical vector based on the Luttinger-Tisza approach~\cite{Luttinger1946}. In reciprocal space, the orbital exchange Hamiltonian in the spinor basis $\bm{\tau}_{\bm{k}}=\left[
	\tau_{z\mathcal{A}}^{\bm{k}},\tau_{x\mathcal{A}}^{\bm{k}},\tau_{y\mathcal{A}}^{\bm{k}},
	\tau_{z\mathcal{B}}^{\bm{k}},\tau_{x\mathcal{B}}^{\bm{k}},\tau_{y\mathcal{B}}^{\bm{k}}\right]$ takes the following form
	\begin{equation}
		H_\text{OE}=
		\sum_{\bm{k}}
		\bm{\tau}_{\bm{k}}^\dagger\Lambda_{\bm{k}}\bm{\tau}_{\bm{k}},
		\Lambda_{\bm{k}}=
		\left[
		\begin{matrix}
			0 & \mathbb{J}_{\bf k} \\
			\mathbb{J}^\dagger_{\bf k}  & 0
		\end{matrix}
		\right],
		\label{eq:HOEKS}
	\end{equation} 
	with
	\begin{equation}
		\mathbb{J}_{\bm{k}}=
		\left[
			\begin{matrix}
				\left(\frac{1}{4}J+J^\prime\right)\lambda^+_{\bm{k}}+J+J^\prime
			 	& \frac{\sqrt{3}}{4}J\lambda^-_{\bm{k}} 
			 	&	0	\\
				\frac{\sqrt{3}}{4}J\lambda^-_{\bm{k}}
			 	& \left(\frac{3}{4}J+J^\prime\right)\lambda^+_{\bm{k}}+J^\prime
			 	&	0	\\
				0  
				& 0 
				& J^\prime\left(\lambda^+_{\bm{k}}+1\right)
			\end{matrix}
		\right]
	\end{equation}
	Here, the dispersion $\lambda_{\bm{k}}^\pm=\exp\left[ik_1\right]\pm\exp\left[ik_2\right]\}$ with the crystal momentum $k_{i=1,2}=\bm{k}\cdot\bm{a}_i$ being expressed in terms of the primitive vectors $\{\bm{a}_1,\bm{a}_2\}$. The orbital interaction can be minimized via the diagonalization of the matrix $\Lambda_{\bm{k}}$. As shown in Fig. 5(a) of main text, the lowest eigenvalues of $\Lambda_{\bm{k}}$ are found to be twofold degenerate at $\pm K$ valleys. The explicit form are as follow
	\begin{equation}
		\psi_{\pm K} = \frac{1}{2}\left[1,\pm i, 0, -1,\pm i, 0\right]^\text{T}.
	\end{equation}
	The coherent superposition of these Bloch waves at $\pm K$ valleys
	\begin{equation}
		\Psi\left(\bm{r}\right)=\frac{1}{\sqrt{2}}
		\left[e^{+i\varphi/2}\psi_{+K}\left(\bm{r}\right)
		+e^{-i\varphi/2}\psi_{-K}\left(\bm{r}\right)\right]
		\label{eq:cgs-s}
	\end{equation} 
	ensures the real-valued pseudospin $\bm{\tau}$ by preserving the time-reversal $\mathcal{T}$ symmetry. The ordered orbital pseudospin in Eq.~(\ref{eq:cgs-s}) is characterized by the azimuth angle $\phi$ in $zx$ plane as  
	\begin{equation}
		\bm{\tau}\left(\bm{r}\right)=\cos\phi\tau_z+\sin\phi\tau_x
		\text{ with }
		\phi\left(\bm{r}\right) = 
		\left\{
		\begin{matrix}
			-K\cdot\bm{r}-\varphi/2 
			&	\text{for sublattice } \mathcal{A}	\\
			K\cdot\bm{r}+\varphi/2+\pi
			& 	\text{for sublattice } \mathcal{B}
		\end{matrix}
		\right.
		.
	\end{equation} 
	As a result, the inference of Bloch waves at $\pm K$ valleys triples the Wigner-Seitz cell of honeycomb lattice. We remark that the Hamiltonian of Eq.~(\ref{eq:HOEKS}) in absence of $J^\prime$ terms is geometrically frustrated with a macroscopic degeneracy of the classical ground-state manifold~\cite{Nagano2007,Nasu2008,Zhao2008,Wu2008b}. 
	
	\section{Linear orbital wave}
	\label{app:LOW}
	Having established the classical ground state of the orbital exchange Hamiltonian $H_\text{OE}$ in Eq.~(\ref{eq:cgs-s}), we then proceed to derive the Hamiltonian that describes the orbital excitation within the linear spin wave theory~\cite{Holstein1940}. In order to study the anisotropic orbital interaction, it is convenient to transform to a rotating frame with the $z$ axis pointing along the pseudospin ordered direction. On each sublattice of the tripled Wigner-Seitz cell, we may rotate the orbital pseudospin
	\begin{equation}
		T_\alpha\left(\bm{r}\right)=\sum_\mu \hat{e}^\mu_\alpha\left(\bm{r}\right)\tau_\mu\left(\bm{r}\right),
	\end{equation}
	where the local frame is spanned by the orthonormal triad
	\begin{equation}
		\hat{e}_\alpha\left(\bm{r}\right)\times\hat{e}_\beta\left(\bm{r}\right)
		=\epsilon_{\alpha\beta\gamma}\hat{e}_\gamma\left(\bm{r}\right).
	\end{equation}
	The explicit relation between local and global frames can be simply written as a pseudospin rotation about $y$ axis
	\begin{equation}
		\left[
		\begin{matrix}
			T_z \\
			T_x  
		\end{matrix}
		\right]
		=
		\left[
		\begin{matrix}
			\cos\phi & \sin\phi \\
			-\sin\phi  & \cos\phi
		\end{matrix}
		\right]
		\left[
		\begin{matrix}
			\tau_z \\
			\tau_x  
		\end{matrix}
		\right].
		\label{eq:local-s}
	\end{equation}
	The local orbital pseudospin operators, obeying the angular momentum algebra of spin $T=1/2$, can be expressed in terms of Holstein-Primakoff bosons~\cite{Holstein1940}
	\begin{subequations}
		\begin{eqnarray}
			T_x&=&\sqrt{\frac{T}{2}}\left(a^\dagger+a\right), \\
			T_y&=&i\sqrt{\frac{T}{2}}\left(a^\dagger-a\right), \\	
			T_z&=&T-a^\dagger a.
		\end{eqnarray}	
	\end{subequations}
	The pseudospin in Eq.~(\ref{eq:OES}) is first replaced by the above transformation.
	Expanding in powers of $T$ followed by the Fourier transformation then leads to the $\varphi$-dependent Hamiltonian in terms of Holstein-Primakoff bosons. To illustrate, we explain the orbital exchange along a generic bond $\bm{e}=\cos\theta\hat{x}+\sin\theta\hat{y}$ connecting $i$-th and $j$-th sublattices as an example. The Hamiltonian with quadratic order in $T$ that accounts for the classical ground-state energy reads
	\begin{equation}
		E_\text{CGS} = 
		JT^2\cos\left(2\theta-\phi_i\right)\cos\left(2\theta-\phi_j\right)
		+J^\prime T^2\cos\left(\phi_i-\phi_j\right).
	\end{equation}
	Moreover, the Hamiltonian with the order $T^{3/2}$ vanishes identically due to the minimization of orbital interaction after summing over all bonds. Importantly, the linear-order Hamiltonian that describes the orbital fluctuation is given by
	\begin{eqnarray}
		H_\text{LOW} =
		&J&T\left[
		-\alpha
		\left(a^\dagger_ia_i+a^\dagger_ja_j\right)
		+\beta\left(a^\dagger_ia^\dagger_j+a^\dagger_ia_j+a_ia^\dagger_j+a_ia_j\right)
		\right]\nonumber\\
		+&J^\prime& T\left[
		-\alpha^\prime
		\left(a^\dagger_ia_i+a^\dagger_ja_j\right)
		+\beta^\prime_-
		\left(a^\dagger_ia^\dagger_j+a_ia_j\right)
		+\beta^\prime_+
		\left(a^\dagger_ia_j+a_ia^\dagger_j\right)
		\right]
	\end{eqnarray}
	with the auxiliary functions $\{\alpha,\beta\}=\{\cos\left(2\theta-\phi_i\right)\cos\left(2\theta-\phi_j\right),
		\frac{1}{2}\sin\left(2\theta-\phi_i\right)\sin\left(2\theta-\phi_j\right)\}$
	and $\{\alpha^\prime,\beta^\prime_{\pm}\}=\{\cos\left(\phi_i-\phi_j\right),
		\frac{1}{2}\left[\cos\left(\phi_i-\phi_j\right)\pm1\right]\}$.
	After the Fourier transformation, we arrive at the following Bogolubov-de Gennes form	
	\begin{equation}
		H_\text{LOW} =
		JT\sum_{\bm{k}}
		\left\{
		\left[
		\begin{matrix}
			-\alpha & \beta_{-\bm{k}} & 	0	&	\beta_{-\bm{k}}\\
			\beta_{\bm{k}}  & -\alpha &	\beta_{\bm{k}}	&	0\\
			0 & \beta_{-\bm{k}} & 	-\alpha	&	\beta_{-\bm{k}}\\
			\beta_{\bm{k}}  & 0 &	\beta_{\bm{k}}	&	-\alpha\\
		\end{matrix}
		\right]
		+2\alpha
		\right\}
		+J^\prime T\sum_{\bm{k}}
		\left\{
		\left[
		\begin{matrix}
			-\alpha^\prime & \beta^\prime_{+,-\bm{k}} & 	0	&	\beta^\prime_{-,-\bm{k}}\\
			\beta^\prime_{+,\bm{k}}  & -\alpha^\prime &	\beta^\prime_{-,\bm{k}}	&	0\\
			0 & \beta^\prime_{-,-\bm{k}} & 	-\alpha^\prime	&	\beta^\prime_{+,-\bm{k}}\\
			\beta^\prime_{-,\bm{k}}  & 0 &	\beta^\prime_{+,\bm{k}}	&	-\alpha^\prime\\
		\end{matrix}
		\right]
		+2\alpha^\prime
		\right\}
	\end{equation} 
	in the spinor basis $\left[a_{i\bm{k}},a_{j\bm{k}},a^\dagger_{i-\bm{k}},a^\dagger_{j-\bm{k}}\right]$. Here the coefficients $\beta_{\bm{k}}$ and $\beta_{\pm,\bm{k}}$ are the Fourier transformation of $\beta$ and $\beta_{\pm}$, respectively. The linear-order Hamiltonian can be diagonalized via the Bogoliubov transformation~\cite{Blaizot1986}
	\begin{eqnarray}
		\left[
		\begin{matrix}
			a_{\bm{k}} \\
			a^\dagger_{-\bm{k}}  
		\end{matrix}
		\right]	
		=
		T_{\bm{k}}
		\left[
		\begin{matrix}
			b_{\bm{k}} \\
			b^\dagger_{-\bm{k}}  
		\end{matrix}
		\right], 		
	\end{eqnarray}
	which relates the Holstein-Primakoff bosons with orbital excitation modes. 
	Notably, the bosonic statistics require that $T_{\bm{k}}$ satisfies the para-unitary condition
	\begin{eqnarray}
		T^\dagger_{\bm{k}}\Sigma_zT_{\bm{k}}=\Sigma_z 
		\text{ with } \Sigma_z=
				\left[
		\begin{matrix}
			\mathbb{I}	&	0 \\
			0	&-\mathbb{I}  
		\end{matrix}
		\right]	.
	\end{eqnarray} 
	Therefore, the linear-orbital-wave Hamiltonian can be written in terms of diagonalized bosons 
	\begin{eqnarray}
		H_\text{LOW}\left(\varphi\right)=
		\sum_{i\bm{k}}2E_{i\bm{k}}\left(\varphi\right)b^\dagger_{i\bm{k}}b_{i\bm{k}}
		+NE_{\text{ZP}}\left(\varphi\right)
	\end{eqnarray}
	where the energy from zero-point motion per unit cell is 
	\begin{eqnarray}
		E_{\text{ZP}}\left(\varphi\right)=
		\frac{1}{N}\sum_{i\bm{k}}E_{i\bm{k}}\left(\varphi\right)-\frac{9}{2}J.
	\end{eqnarray}
	Finally, the zero-point energy can be evaluated numerically.

\end{document}